\documentclass[11pt]{article}
\usepackage{graphicx,amsmath,amsfonts,amssymb,dcolumn,amsthm,euscript}
\usepackage{tikz}
\usetikzlibrary{snakes}
\usepackage{color}

\newcommand\nn{\nonumber}
\newcommand\eq{\end{equation}}
\newcommand\bq{\begin{equation}}

\def\bar{\overline}
\newcommand\pa{\partial}
\newcommand\bra{\langle\!\langle}
\newcommand\ket{\rangle\!\rangle}
\def\hb{\hbar}
\def\d{{\rm d}}
\newcommand\FP{\mathrm{FP}}
\newcommand\cl{\mathrm{cl}}
\newcommand\susy{\mathrm{susy}}
\newcommand\stoc{\mathrm{stoc}}
\newcommand\topo{\mathrm{top\,}}

\newcommand\del\delta
\newcommand\lam\lambda
\newcommand\Del\varDelta
\newcommand\Ph\varPhi
\newcommand\Ps\varPsi

\newcommand\mint[2]{\raisebox{.1ex}{\scalebox{0.85}[0.85]
 {$\displaystyle\int_{#1}^{#2}$}}}
\newcommand\sfrac[2]{\raisebox{-.1ex}{\scalebox{1.1}[1.1]{$\frac{#1}{#2}$}}}
\newcommand\mfrac[2]{\raisebox{-.1ex}{\scalebox{1.2}[1.3]{$\frac{#1}{#2}$}}}

\begin{document}
\title{\textbf{Second Order Langevin Equation and Definition\\
of Quantum Gravity By Stochastic Quantisation}}
\author{\textbf{Laurent~Baulieu$^\dagger$ and Siye~Wu$^{\dagger\dagger}$}
\thanks{{\tt baulieu@lpthe.jussieu.fr},\quad{\tt swu@math.nthu.edu.tw}}\\\\
\textit{$^\dagger$ LPTHE, Sorbonne Universit\'e, CNRS} \\
\textit{{} 4 Place Jussieu, 75005 Paris, France}\\
\textit{{$^{\dagger\dagger}$Department of Mathematics,
National Tsing Hua University}}\\
\textit{{30013 Hsinchu, Taiwan}}}

\date{}
\maketitle
\begin{abstract}
Euclidean quantum gravity might be defined by stochastic quantisation that
is governed by a higher order Langevin equation rather than a first order
stochastic equation.
In a transitory phase where the Lorentz time cannot be defined, the parameter
that orders the evolution of quantum gravity phenomena is the stochastic time.
This changes the definition of causality in the period of primordial
cosmology.
The prediction of stochastically quantised gravity is that there will a
transition from an oscillating quantum phase to a semi-classical one, when
the Lorentz time emerges.
The end of the transition, as it can be observed from now and described by
inflation models, is a diluted Universe, following the inflation
phenomenological evolution.
It is filled at the beginning with scattered classical primordial black
holes.
The smallest ones will quickly decay in matter, with a standard quantum
field theory evolution till our period.
The stable heavier black holes will remain, forming a good fraction of the
dark matter and the large black holes observed in the galaxies.
In a theoretically related way, this framework suggests the possibility of
a gravitational parton content for ``point-like" particles, in the same
five dimensional quantum field theory context as in the primordial cosmology,
with a $(+----)$ signature for the $5$d metrics.
The very precise and explicit result expressed in this paper is actually
far more modest than its motivation.
We compute explicitly the meaning of a second order Langevin equation in
zero dimensions and define precisely what is second order stochastic
quantisation in a soluble case.
\end{abstract}

\newpage

\section{Introduction}
In has been proposed in \cite{cosmob} that $4$d Euclidean quantum gravity
might obey the laws of a stochastic quantisation governed by a second order
rather than a first order stochastic equation and, perhaps equivalently, by
a possibly higher order Fokker-Planck equation and its $5$d supersymmetric
representation.

The idea of \cite{cosmob} is that in the case of gravity, this equilibrium
can only be the classical gravity theory (coupled to ordinary standard
particle QFT's), due to a given fluctuation that suddenly increases the volume
of the space, with a brutal decrease of the cosmological constant, so that
there is no more room for quantum gravity effects, and Lorentz time can be
then defined.
In this way one avoids the usual paradox that the Lorentz time cannot be
defined when non-perturbative quantum gravity prevails.
The system has two phases, one without Lorentz time and other one that has
an ``emerging'' Lorentz time, but both phases rely on the same microscopic
theory.
It was observed that this may enlarge the scope of causality.

In this paper we explain in more details how stochastic quantisation can
truly fits the proposition of~\cite{cosmob} and find a modification of the
theory that explains the exit from inflation.
There is no contradiction between our proposition and the present
understanding of various aspects of semi-classical gravity in the
post-inflation regime.

Many aspects of the Langevin equation we propose here are in a separate and
more mathematically oriented paper~\cite{bcw}, published elsewhere.
The purpose of the present paper is to introduce and justify physically a
second order Langevin equation as a method of quantising gravity.

More importantly, we predict some physical consequences in cosmology and on
the short distance behaviour of point-like particles.
Some remarks will be also given to support the use of a second order Langevin
equation for non-perturbative gravity.
Stochastic quantisation in Euclidean quantum gravity is in fact an
out-of-equilibrium system from the point of view of the Euclidean path
integral.
It is indeed well known that for $d>2$ the Euclidean Einstein action
$S_{\mathrm{E}}$ is not bounded from below and hence its exponential
$\exp(-\frac1\hbar S_{\mathrm{E}})$ is not suitable for defining an
equilibrium distribution. 

In this paper we argue that the modification from the first order stochastic
quantisation Langevin equation of Paris-Wu \cite{parisi} to a second order
one opens new possibilities during the relaxation phenomenon toward the
quantum physics equilibrium.
We expose the following claim: physically relevant fluctuations of the
vacuum of the $5$-dimensional supersymmetric quantum field theory that can
be associated to any given Langevin equation from a $4$d Euclidean quantum
field theory may occur at finite stochastic time, in correlation with quantum
effects consisting of an abundance of creations and annihilations of the
quantised states of the systems.
In the case of the stochastic quantisation of gravity, we call them
$5$-dimensional gravitational partons, in analogy with the physical scheme
discussed in \cite{cosmob}.

In the case of gravity, stochastic time fluctuations of the metric field change
the structure of the space when the stochastic time evolves.
The change in geometry can be so strong that it can be assimilated to a reset
of the initial conditions, putting the system in a very different phase. 
This supports our view that they can trigger a transition from a regime where
quantum gravity must be treated in a completely non-perturbative way, that is
by the $5$d stochastic quantisation method, to a semi-classical regime where
it can be treated by the standard $4$d perturbative methods.

Using a standard first order stochastic quantisation is actually a dead end
in the case of gravity because it has no equilibrium distribution and its
dynamics is ill-defined.
But if one uses higher order stochastic quantisation, new phenomena may occur
in finite stochastic time and they may help avoid the known difficulties
encountered in the standard attempts of quantum gravity.
The creation of pairs of microscopic $5$d quantised states is perhaps the
cause of a possible phase transition at any given moment, explaining the exit
from the primordial cosmology phase towards the phase where the universe is so
diluted that it can be described by the standard quantum gauge field theories
in a (semi)classical gravitational background.
(At a much more modest but rather realistic level, some progress in the theory
of earthquakes was achieved by using models with a second order Langevin
equation rather than a first order one.) 

Our picture for quantum gravity is therefore a very non-trivial generalised
Brownian motion.
Namely, it has a phase transition, marked by the exit from the inflation, the
meaning of which will be made precise below.
The occurrence of such oscillations is made rather explicitly (in the
Appendices) in a zero-dimensional interacting model, treated perturbatively.
In higher dimensions, counterbalancing between the oscillations of the vacuum
and the productions and absorptions of partons may occur during the stochastic
time evolution.
The zero-dimensional case is of course simplistic in comparison to the one we
propose for quantum gravity in four dimensions.

The second order Langevin equation that we propose here has a drift force for
the metric that is the sum of the Einstein tensor (plus a possible contribution
proportional to the cosmological constant) and the energy-momentum tensor of
the matter.
An additional gauge-fixing drift force takes into account the $4$d
reparametrisation invariance that must hold true at every given finite value
of the stochastic time.
The result is a very interesting and complicated differential equation
involving a noise whose effect vanishes when $\hbar\to0$.
It generalises the flow equation of classical gravity within the scheme of
stochastic quantisation by addition of a noise (multiplied by $\sqrt\hbar$)
to the gravity drift forces.
Some mathematical properties of this equation will be explored in \cite{bcw}.

Stochastic quantisation is intrinsically set up for defining the quantum
Euclidean correlators of a theory classically defined by its equations of
motion.
It provides a wider perspective and deeper insights to the signification of
the path integral method \cite{parisi}.
Given a drift force and certain initial conditions, it is logically possible
not to be able to Wick rotate one of the Euclidean coordinates in the
correlators that stochastic quantisation determines at a finite stochastic
time.
If the stochastic time is the true evolution parameter of physics, the Lorentz
time simply doesn't exist during the phase of primordial cosmology.
In this phase the Physics must be described in the $5$-dimensional framework,
just as one must use the microscopic time as the parameter of evolution when
describing a Brownian motion far enough from its equilibrium.
In fact, it would make no sense to have the coexistence of both the
stochastic time and the Lorentz time for ordering phenomena from the knowledge
of correlators.
No theory can be defined with two independent parameters of evolution.

This point is actually an advantage in gravity theory, in which we know that
the Lorentz time cannot be part of the hypothesis in the definition of quantum
gravity beyond perturbation theory.
This is for instance obvious because of the contradictions 
implied by the Wheeler-deWitt equation.
By giving a physical meaning to the evolution in the stochastic time, our aim
is in fact to obtain a substitute to order physical events before the
transition identified by the exit of the inflation.
So, given the impossibility of defining a Lorentz time when gravity has a
non-perturbative quantum behaviour, we propose that the parameter that could
truly describes the evolution of processes is the stochastic time.
This results in a broader sense of causality in the primordial phase of the
Universe, made explicit by the Langevin equation.

The usual description of causality in the evolution in Lorentz time is of
course recovered after the exit from inflation when the Lorentz time can
emerge and gravity can be described semi-classically. 
After this event, all the correlators of the fields can be computed directly
by the standard path integral in the smooth and existing limit of infinite
stochastic time.
Their analytic continuation is possible, as a general result of perturbation
theory, allowing the emerging Lorentz time to be used after Wick rotation as
a phenomenological ordering parameter.
It is only in this phase that quantum field theory predicts the existence of
clocks built on the principle of using the computable decay rates of atoms or
elementary particles.
Notice that the determination of such rates only involves correlators that
can be directly computed in the purely Euclidean theory.

This fancy way of defining the Lorentz time (when it can possibly exist)
can be also applied to classical physics.
Indeed, one can replace the Hamilton equations of motion by flow equations.
This may look as a formal remark, but the case of Ricci-flow equations in
gravity is still an open question.

A conclusion of the heuristic scenario of \cite{cosmob} was that the
interpretation of the Euclidean correlators at a finite stochastic time is what
might define quantum gravity until one has a sudden relaxation to semi-classical
gravity by a phase transition identified as the exit from inflation.
The second order Langevin equation we present here make this scenario
plausible in a more precise way.
More mathematical details on the Langevin equation than what we will exhibit
here will be given in \cite{bcw}.

We are aware that the widely discussed question of defining time in quantum
gravity, the so-called ``time problem" has been approached in many different
ways.
We will not refer to all these other approaches, because ours predicts that
in the presence of non-perturbatively gravity there cannot be a possible
definition of the Lorentz time.
We believe therefore that there is not much sense to compare the content of
this paper and that of other interesting articles about the question of the
Lorentz time.
Rather, we should only be concerned with the compatibility of our proposition
and the already established facts in (semi-classical) quantum gravity, and we
will discuss these topics in the paper. 

Besides, the quantum gravity community is often unfamiliar with the possible
improvements brought to the quantisation methods by the use stochastic
quantisation.
Part of our intuition relies on a comparison between the the spin-glass
theory that cannot correspond to a regular Boltzmann thermalisation and
quantum gravity that cannot undergo a smooth infinite stochastic limit in
the framework of stochastic quantisation. 

Let us explain why.
Spin-glass are systems with an out of thermal equilibrium phase that can be
reversibly melted into an ordinary phase with a regular thermal distribution. 
There is an analogy between the ``phase transition'' for the stochastic
stochastic time evolution of gravity and what happens to spin glasses.
In the spin glass phases, thermal equilibrium cannot be reached.
Their distinct physics cannot be described by Gibbs equilibrium but depends
on the initial conditions, and their causality properties are quite special,
with with unusual evolutions and no stable equilibrium.
However, at the microscopic level, spin glasses are made of standard matter.
Thus they can undergo phase transitions, like melting or sublimation, in which
the spin glass becomes a standard liquid or gas, whose properties can be
described by a regular Boltzmann-Gibbs ensemble.
In this way, the system can evolves toward one in an ordinary thermal
equilibrium, under the influence of changes of external parameters.
Technical advances have been done to compute the correlators of the out of
thermal equilibrium phase of spin glass and describe their physics by using
Langevin equations.

For gravity, the transition between its non-perturbative phase, without
Lorentz time and standard path integral $4$d formulation, and its
semi-classical gravity phase might exist in a way similar to the possible
transition for spin-glasses.
In each case, both phases are described by the same Langevin equation of
the same microscopic theory.

The non-perturbative phase of gravity is a non smoothly converging regime
of pure fluctuations induced by the noise.
There is no way out other than using a Langevin equation do compute the
correlators, which give a different physics.
For the semi-classical phase, the techniques of computations greatly simplify
by using the path integral formula (as one uses the Boltzmann formula to
describe a melted spin glass)\footnote{Part of the comparison is that the
stochastic time is like the physical molecule time for the collisions of
particles in a Brownian motion, and the Lorentz time, when it can be defined,
is like the adiabatic time of measurements when one measures the effects of
the variation of external parameters for a system made of a very large number
of the molecules in thermal equilibrium.}.
Notice that in classical physics, the use of flow equations rather than
equations of motion is the same as using a Langevin equation while
neglecting the noise. 

A tempting idea is therefore to compare of impossibility of non-perturbative
quantum gravity to reach its infinite stochastic time limit ``smoothly" i.e.,
without a sharp transition) with the impossibility of reaching a standard
thermal equilibrium in the spin glass example.
This is modulo the fact that the transition between the ``phases'' of quantum
and classical gravity is ``spontaneous''.
Here, spontaneity means that the effect of a single fluctuation, without a
change in external parameters, can trigger the change of phase, giving a
possible very sharp transition.
It can be so because the fluctuations of the gravitational field under the
effect of the stochastic noise affect the structure of the space.
In return, this effect can trigger a transition toward a phase where the
system becomes basically semi-classical everywhere, as if the stochastic
process had been brutally reset with boundary conditions compatible to a
semi-classical description of the system.

In this phase, analogous to that of a melted spin-glass with a consistent
Gibbs distribution, the Euclidean correlators admit a well defined and
smooth large stochastic time limit, in which case their analytic properties
are such that the Wick rotation is possible at very large stochastic time,
a phenomenon that we called the ``emergence'' of Lorentz time.
An example where ergodicity is spontaneously recovered by a phase transition
with no need of changing the parameters is the decoherence of a microscopic
quantum system, coupled to a larger classical system. 

The comparison with spin glasses has some limitations.
The exit from inflation in gravity is not a reversible phenomenon because it
concerns with the whole system and the probability that it can spontaneously
reverse its course is zero.
The transition is spontaneous and induced by the fluctuations of the metric
field, whose induced changes of correlators affect the boundaries of the
system.
In contrast, the transition in a spin glass is not spontaneous.
Its melting must be imposed from the outside, for instance by external
sources that change locally the temperature. In this way, it is reversible.

Let us recall that such a spontaneity scenario was already present of
\cite{cosmob}, and based on a Markov process, with the heuristic idea that
in the primordial cosmology conditions, the time could flow in a discrete way.
Here the interpretation of the oscillation of Euclidean correlators at a
finite stochastic time defines precisely quantum gravity, and also predicts a
possible sudden relaxation to classical gravity by a phase transition
identified as the exit from inflation.

The paper is organised as follows.
Section~2 justifies the use of stochastic time to order non-perturbative
quantum gravity phenomenon.
Section~3 explains stochastic quantisation with a first order Langevin
equation.
Sections~4 displays materials needed for second order Langevin equations.
The use of such equations is illustrated in Appendix~A by an interacting
zero-dimensional example, making concrete the meaning of a second order
Langevin equation (at the perturbative level).
We compute the oscillation mechanism that is the basis of our description of
gravity with a second order Langevin equation.
Appendix~B describes the supersymmetric representation for such second order
Langevin equations.
This gives some hint about the oscillation in stochastic time of a classical
background field around which one defines a quantum theory.
Section~5 gives some justifications for using stochastic quantisation for
gravity. 
Section~6 presents the second order Langevin equation for the gravity.
This equation will be studied in depth in \cite{bcw} by checking how it can
preserve the $4$d reparametrisation invariance throughout the stochastic time
evolution.
Section~7 concludes with further speculations about gravitational phenomena
suggested by the use of a second order Langevin equation for quantum gravity.

\section{The stochastic time as the evolution parameter of primordial
cosmology}
One must be more precise about what we mean for the term ``exit from
inflation'' through this article.
The claim of this paper is that second order stochastic quantisation might be 
a key to define the non-perturbative quantum gravity fluctuations that must
occur in primordial cosmology to reach a regime where gravity can be treated
semi-classically everywhere in the Universe.
In this framework, the exit from inflation period means the end of the
relatively narrow cross-over domain between the stochastically disordered
phase of quantum gravity and the phase of semi-classical gravity.
At the very end of this cross-over, the Lorentz-time has ``approximatively
emerged," but residuals of the non-perturbative gravity quantum regime
persist, in a more and more negligible way, with a continuous evolution
toward a situation where gravity can be treated semi-classically everywhere.
For this end of the transition, one may use the $4$d formalism by an ad hoc
completion of the fundamental fields with a given set of inflaton fields
(perhaps of no fundamental significance), allowing the use of an inflation
model whose evolution is phenomenologically parametrised by a distorted
Lorentz time.
This provides a local quantum field theory that may reliably describes the
physics of the end of the crossover.

The effect of inflaton fields fades away as they are no longer needed in the
final phase where all possible effects of gravity can be treated
semi-perturbatively. 
The current inflation theory scenario indicate that the phenomenological
inflaton fields get stabilised to constant values after some delay.
The cosmological constant gets equal the small constant we measure today and
the semi-classical gravity regime stays irreversibly as it is now, with its
own standard causality laws.

So, in our scheme, the ``exit from inflation'' is marked by the possibility
of expressing the evolution of the physics either in Lorentz or in Euclidean
formulation (equivalently) at infinite value of the stochastic time.
Only afterward can one use the emerging Lorentz time as our privileged
parameter to order the phenomena via a causal and local QFT, with a standard
classical limit.

In the relatively very short duration of the exit of the inflation, there are
some remnants of the non-perturbative effects of quantum gravity.
The dynamics of both phases mixes during the crossover, but the virtue of the
inflation theory is to phenomenologically describe these mixed effects.
This happens till the non-perturbative quantum effect of gravity has
disappeared from the dynamics of the Universe.
Thus, for the special moment of the history of the Universe, using a Lorentz
time is made possible at the cost of using ad hoc and non-fundamental inflaton
fields.

Before the exit of the inflation, the stochastic quantum field theory
framework is necessary to compute the equal finite stochastic time field
correlators of the Euclidean correlators along the lines indicated in
Sections~3 and 4, with some illustrations in Appendices~A and B.
The task is of course far of being understood for the case of gravity and 
we can only describe the method from a generic point of view.

Stochastic quantisation relates fields to their noises.
The second order Langevin equation~\eqref{gra} that will be considered for
gravity is a PDE that relates the metric field to its noise.
When one solves it and computes field correlation functions by averaging
over the noises at finite values of the stochastic time, the result depends 
on the initial field configuration at an earlier stochastic time. 
It follows that the early cosmology evolution will depend strongly on the
new scales introduced by the Euclidean geometry one chooses as an initial
condition for the stochastic process.
The latter is an initial condition in the Cauchy problem of the Langevin
equation.
It is only under the condition that a stochastic equilibrium has been
reached that the values of correlation functions become independent of the
initial conditions.
This property is nothing but the generalisation of what happens in the
Brownian motion of a gas.
Its distribution at finite stochastic time depends on the initial condition.
It is only when the thermal equilibrium is reached is the dependence on the
initial conditions is truly washed away, giving a Boltzmann partition
function.

In the case of a second order Langevin equation, the rate of fluctuations
influences the probability of a possible phase transition on the way to a
thermal equilibrium.
This rate is basically given by the dimensionful constant $\Del T$ that must
occur for dimensional reasons in the acceleration term in a second order
Langevin equation as Eq.~({\ref{gra}).
The smaller $\Del T$ is, the faster one has the possibility of a phase
transition.

Consider now what may happen just after the phase transition we advocate, when
gravity can be computed in its semi-classical approximation.
The scale of variation of all observable physical phenomenon is much (almost
infinitely) slower that the convergence of the stochastic process, as in a
Brownian motion.
One can take the limit of infinite stochastic time, and then the details of
the stochastic time evolution become physically irrelevant for all realistic
experiments.
The dependence on the initial data of the correlation functions in the phase
before the phase transition is in fact completely washed away after its
relaxation to the phase where gravity has become effectively (semi)classical
everywhere, as a typical thermalisation phenomenon, and standard clocks for
the Lorentz time can exist and be located at any given point of the space. 

For instance, such standard clocks, located at various places of our observable
Universe, will never allow us to get information about the initial data that we
might wish to know to solve (in theory) the Langevin equation in the primordial
cosmology phase.
Indeed, when one computes the rate of physical Lorentz time clocks in the
semi-classical regime of gravity, there is no need to be concerned with the
details of stochastic quantisation. 

However, the understanding of the stochastic quantisation scheme can help
figure out an appealing set of initial conditions right after the exit from
inflation, giving a more theoretically sound framework to the physical
description of \cite{cosmob}.
It is interesting to compare the use of a Langevin equation with the discrete
time scenario of primordial cosmology.
As discussed in \cite{cosmob}, there is a possible justification for having a
sharp exit from the inflation by the hypothesis that the Lorentz time cannot be
considered as a continuous parameter at very short distance and might run in
a discrete way during the primordial cosmology.
In the discrete time framework, $\Del T$ is the interval between two successive
steps of the discrete time, while in this paper, the scale $\Del T$ is directly
introduced as the coefficient of the second order term in the Langevin equation
and the four Euclidian coordinates $X^\mu$ are continuous. 
This suggests the possibility of interpreting the space with coordinates
$\{X^\mu,\tau\}$ as a $5$d-space foliated by the stochastic time parameter
$\tau$.
This will be done separately in the more mathematically motivated work
\cite{bcw}.

A striking feature of the stochastic time picture is that it justifies quite
logically the exit from inflation as a phase transition, allowing the Lorentz
time to emerge at an age of around $10^{-30}s\gg\Del T$ (or perhaps less,
provided it is much bigger than $\Del T$), which is about the limit of how far
we observe our past.
In this way, when it can emerge, the Lorentz time is nothing but one of the
Wick rotated coordinates $X^\mu$, in complete agreement with the usual wisdom.
This is quite a clarification compared to \cite{cosmob}.

Only after this transition, when all correlators can be computed by the 
conventional $4$d path integral methods, do the standard QFT properties imply
analytic behaviours of the correlators, allowing their inverse Wick rotation. 
Non-perturbative gravitational effects can be neglected everywhere.
At this stage of the Universe, the Lorentz time has basically emerged and
can be used to compute the evolution of the physics in the framework of the
current phenomenological theories.
The recently created Lorentz time has then become a convenient parameter to
order the relative successions of events, as they can be experimentally
observed and precisely measured.

Therefore, an appealing novelty of the approach is that it doesn't postulates
the existence of the Lorentz time.
Rather, it considers the Lorentz time as a possibly measurable coordinate,
which has been analytically continued.
This can only be done after the exit of the inflation in the context of
stochastically quantised Euclidean quantum field theory.
After that one can show the consistency of using the Lorentz as an evolution
parameter both classically and quantumly, because gravity became
(semi) classical, and clocks can exist in this phase.

The existence of the Lorentz time can (and must) be checked of the
possibility of building and using clocks, e.g., made from the created
particles.
Let us stress that in our framework, the theory of the inflation becomes then
a phenomenological model for the short period during which the existence of
the Lorentz time is only ``approximately'' valid, in the sense that shortly
after it the stochastic time dependance becomes undetectable by our
observations, and one can use the $4$d formalism for all physical fields.

After the transition, the Lorentz time can be used effectively sufficiently
far away from the created primordial black holes singularities.
In fact, even before the inflation period is completely over, the inflation 
phenomenological scheme is defined by a ``time dependent cosmological
constant'' that quickly evolves towards a constant value in an irreversible
process.
It provides a local quantum field theory that is phenomenologically admissible
for a ``gracious exit'' from inflation using appropriate effective (unphysical)
inflaton fields, with an effective dependence on the Lorentz time and no
contradiction with the CMB experimental observations.
In \cite{cosmob} we estimated the frequency of the stochastic oscillations
needed to reach a transition from primordial cosmology to our phase in which
semi-classical gravity prevails to be of the order of
$1/\Del T\sim10^{65}s^{-1}$ or more.
This transition necessarily occurs with probability $1$ after a large enough 
enough number of sequences.

The oscillatory regime of stochastic quantisation suggests that the phase
transition that allows the emergence of Lorentz time is marked by a very
large production of gravitational bound states enhanced by the vacuum
fluctuations in the 5d theory\footnote{See Appendix~B for some
zero-dimensional example as a (partial) justification of this statement.}.
Such predictions were heuristically justify in the framework of \cite{cosmob}.
At most, \cite{cosmob}, identified the exit of the inflation as the period
after which one can approximate the Lorentz time as a continuous variable,
so that all quantum correlators become independent of the scale variable
$\Del T$.
Its occurrence was justified by the energy fluctuations implied by the
existence of a discrete time.

The scheme of stochastic quantisation scheme may indicate more precisely
that, as soon as strong enough stochastic fluctuations occur, creation of
an abundance of classical gravitational bound states at a given value of
the stochastic time is what makes the space description semi-classical
everywhere, except at the $4$d singular points of the leftover primordial
black holes, which are scattered everywhere after the exit from inflation.
The discussion in the Appendix~B is a modest step in this direction.

So we propose in this paper that, as a consequence of the description of
primordial Cosmology by the $5$d stochastically quantised gravity, the
initial condition after the exit of inflation of the phenomenological
inflation equations is a small universe, filled with a cloud of primordial
black holes with a given mass distribution. 
The latters are remnants of the $5$d-gravitational bound states on the
$4$d boundary at infinite stochastic time (our observable Universe).

This suggestive picture describes a bit more precisely than in \cite{cosmob}
the phase transition from quantum to classical gravity and we claim that it
is a genuine consequence of having a non-perturbative quantum gravity regime
driven by an evolution in the stochastic time defined by a second order
Langevin equation.

It predicts that whatever the exact shape of the end of inflation period is,
it drives the Universe towards its semi-classical gravity phase where its
$4$d geometry is determined in an excellent approximation almost everywhere
by solving the Einstein equation with a fixed cosmological constant with
appropriate boundary conditions.
The relatively short period in which an effective theory exists with a Lorentz
time only approximatively true is a necessary feature.
It makes a smooth extrapolation between the end of the purely quantum gravity
regime and the beginning of the semi-classical regime of our phase.
This semi-classical regime means the geometry can be treated classically at
an excellent approximations modulo the possible emissions of perturbative
gravitons, all other interactions being defined by their standard $4$d
quantum field description. 
No realistic experiments can detect in our phase the non-perturbative quantum
gravity properties, although its Langevin equation is the same as in the
primordial phase.

The detailed evolution at the very end of the inflation is just whatever it
was, and is not so meaningful, nor is the exact shape of the curve
describing the end of a cross-over region.
It is in fact well known in other branches of physics that, for most phase
transitions, the details of the precise cross-over curve are very non-trivial
as they depend on initial conditions which are in general not under control.
As a consequence, the period following the exit from inflation can occur with
or without reheating, that is, with or without oscillations in the recently
created Lorentz time.
There is a wealth of inflation models that can describe all possibilities.
Such details are phenomenologically important, although they occur in a
relatively short duration of perhaps $\sim 10^{-33}$s. 
They are of great help as their signals seem to be almost certainly
observable from precise CMB measurements, which is a great success of the
inflation model.

So, at this level, the use of stochastic quantisation as the quantum
formalism for all interactions is consistent with the description of the
last moment of the exit from inflation by a standard inflation theory.
More importantly, our considerations predict a phase transition and the
definition of initial conditions: our phase starts with a gas of primordial
black holes, some of which would shortly decay towards standard matter,
depending on their sizes, and possibly to more exotic particles, for
instance the so-called WIMPS.
The heavier black holes are stable at the time scale of the age of the
Universe, and will remain as a substantial contribution to dark matter, in
addition to what possibly comes from the many post-inflation scenarios.

Our scenario leaves undecided the mysterious question of whether or not a
small black hole can really fully shrink to a state with no black hole at
the end of the process.

The decay of primordial black holes can only occur when the expanded Universe
is large enough compared with the dimension of particles for the latter to be
possibly created.
When this has happened a reliable evolution can be computed according to the
laws of the numerous viable gauge models where gravity is treated
semi-classically.
In this phase the Lorentz time has emerged and it can be used to order
observed physical events.

It is gratifying that this scheme provides a good explanation of the new ideas
that galaxies might often be built around black holes.
Indeed, as it has become obvious, many of these black holes cannot be
obtained by the collapse of standard matter.
We predict here that a large part of the standard matter was created by the
decay of some of these early primordial black holes (e.g., by Hawking like
effects) determined by the transition from quantum to classical gravity, while
the standard black holes stemming from gravitational collapses of stars are
perhaps in the minority.
They are in fact of secondary importance.
Later in the paper, we will also comment on a possible application of this
$5$d framework to giving a gravitational structure to the $4$d point-like
particles, providing meaning to their ultra short range properties.

To summarise, in the framework of stochastic quantisation of quantum gravity
and cosmology, the presence of ``primordial black holes'' is a signature of
an unavoidable stochastic time fluctuation that has triggered the existence
of inflation.
In the post-inflation regime, when the Universe is big enough for gravity
to be classical and small enough for the standard matter not to exist yet,
it is plausible that the early cosmology can be well approximated by a
scattered cloud of primordial black holes, with a mass distribution that
determines their further decay to standard matter within a background of dark
matter that is the remnant of the heaviest primordial black holes.
The advantage of using stochastic quantisation with a second order term is
that the same equation gives a possibly adequate new physics for the primordial
cosmology and reproduces all known results for post inflation physics.

\section{Answers to possible counterarguments against the use of the stochastic
quantisation scheme of gravity } 
More can be said on whether any quantum behaviour that cannot be described by
a stochastic quantisation equilibrium has consequences consistent with the
known results in semi-classical gravity.

A quite simple possible counterargument can be made against the non-existence
of the Lorentz time in the primordial cosmology.
More precisely, since time is emergent in our picture, it is unclear at the
first sight if one can eventually end with a system which is locally Lorentz
invariant.
In fact, the Lorentz symmetry must be already rooted in the bulk of stochastic
quantisation, although it has no room there, since the Wick rotation cannot be
done at finite stochastic time.
This is not a minor issue because there are very strict bounds on violations of
Lorentz symmetry.

The answer is in \cite{bcw}, which shows that the gravity Langevin equation
computes the evolution of leafs, with metrics $g_{\mu\nu}(\tau,X)$ and
imposes reparametrisation invariance in the Euclidean coordinate coordinate
$x^\mu$ in each leaf at constant $\tau$.
So Eq.~\eqref{gra} is not $5$d invariant, but can be made $X^\mu$ covariant for
each value of $\tau$, in a typical foliation. 
Thus there is a local $SO(4)$ symmetry in each leaf at each value of $\tau$.
This symmetry is preserved in the $\tau$ evolution and survives the exit from
inflation.
So as soon as the stochastic process has converged effectively, one can do the
Wick rotation and an exact emerging local $SO(1,3)$ symmetry is warranted in
our phase. 

Notice that all measurements that relate our epoch to the moment of exit from
inflation have no access to the primordial $5$d physics prior to the exit.
Thus our scheme of primordial cosmology is compatible with all realistically
possible experiments about the strict bounds on violations of Lorentz symmetry.

Notice also that there is no way that we can observe the physics in the phase
of quantum gravity.
Indeed, we have no apparatus to detect the evolution in stochastic time,
which runs at a scale given by the super high frequency $1/\Del T$ of
oscillations in the second order Langevin equation.
All precise information about the early cosmology is washed away after the
exit from inflation analogously as all global information about a spin glass
is erased after it transforms into a standard liquid or solid.

Another subtle point that should be addressed is the following.
According to our model, the exit from inflation terminates the regime of
non-perturbative quantum gravity completely.
The question is whether this is compatible with the arguments for
constraints on the couplings of classical gravity to quantum field theory.

For instance, puzzling questions were presented by Page and Geilker \cite{page}
and in \cite{kent} and references therein, where gedanken (and possibly
realistic) experiments point out the necessity of a mechanism for allowing the
standard collapse of the wave function of quantum matter coupled in a
non-negligible way to classical gravity.
Page and Geilker \cite{page} investigated the consequences of the equation
$E_{\mu\nu}\sim\langle T_{\mu\nu}\rangle$ for the interactions between
classical gravity and quantum matter.
They investigated the necessity for having a wave function collapse of the
matter of which $T_{\mu\nu}$ is the energy-momentum tensor. 
To solve the inconsistencies, it is necessary and probably sufficient that
some perturbative quantum gravity holds after the exit from inflation.
In this way, gaussian perturbative quantum excitation emitted by the classical
gravity field defining the Einstein tensor $E_{\mu\nu}$ will forbid any
possible paradoxes about a violation of the matter wave function collapse.

It is thus necessary that the scheme of stochastic quantisation allows the
exchange of perturbative quantum gravitons in our phase, at least at a very
microscopic level, in order to allow possible back reactions.
The logics is that it must occur as in QED, where analogous wave function
collapse paradoxes disappear when one admits, correctly, that a classical
background can emit or absorb photons. 
So a good question is whether some kind of perturbative gravity truly
survives the exit from inflation, even if its non-perturbative quantum
gravity effects have disappeared effectively.

As a matter of fact, perturbative quantum gravity, although not a
renormalisable theory, is perfectly compatible with stochastic quantisation
at any given finite order of perturbation theory provided the perturbation UV
quantities are suitably regularised\footnote{To get a formal proof of this
fact, one must incorporate in its Langevin equation a gauge restoring force
such as the one in the Langevin equation we will display and justify in a
geometrical way in \cite{bcw}.}.
In fact, what may follow the exit from inflation is the classical gravity
completed by its standard perturbative theory, suitably regularised.
This theory allows emission and absorption of perturbative on-shell gravitons,
much like a classical electric field is able to emit and absorb perturbative
photons.
This possibility is enough to satisfy the first principles of quantum mechanics
including the wave function collapse property in our phase, with experiments
such as those suggested in \cite{page} and \cite{kent}.

Thus, although gravity is basically classical everywhere in our phase, any
gravitational background can be considered as a coherent state rather than
a classical field.
This means that any kind of gravitational background can evolve towards a
slightly different one, modulo the emission or absorption of spin 2 gravitons.
As in the much better understood case of QED, this ensure the possibility of 
back-reactions of a gravitational background on a microscopic quantum system.
This should be sufficient to avoid paradox when one confronts the topics of
measuring the collapse of the matter wave function of quantum matter under the
influence of a gravitational field\footnote{Since we
invoke that gravity may and must involves perturbative effects in our phase,
we may push the question further.
Can we regularise and compensate the ultra-violet divergencies of gravity that
also occur in the stochastic quantisation of gravity by looking at the gravity
theory in a stringy context?
The answer seems positive, since one can for instance treat the string Polyakov
action in a stochastic approach, and it looks as if one can follow
diagrammatically the phenomenon of compensation of ultraviolet divergencies of
gravity in the stochastic quantisation scheme.}. 

Another general question is whether or not one can we fully trust the Hawking
formula of black hole evaporation, which is at best a classical or
semi-classical computation, while neglecting completely aspects of
non-perturbative gravity. 
There is a huge literature about controversies if some remnants of the
black holes should stay when the mass of the black hole reaches values as
small as the Planck mass, or even less.
A better understanding of possible emission of perturbative gravitons 
is probably insufficient for solving this question about the black holes
loss of information.

Stochastic quantisation may or may not bring a modest contribution to these
topics.
On the one hand it provides a new physical cutoff because of the coefficient
$\Del T$ of the second order term of its Langevin equation, and the $\Del T$
dependance might bring new interesting topics.
On the other hand, when matter is sufficiently dilute, no process can carry
enough energy to create conditions as those of the quantum gravity regime of
early cosmology, which require the $5$-dimensional framework.
Thus, the use of stochastic quantisation will probably not help understanding
the fate of a black hole when it shrinks to zero.
We do not know if a combined knowledge of the very short distance of gravity
on the one hand and a better handling of its semi-classical approximation far
away form singularities on the other hand may clarify this issue.

A more positive note is however as follows. 
The Hawking evaporation has some analogy with the Schwinger effect
of matter creation in a strong electric field.
Thus it must involve some quantum back-reaction of the classical gravitation
field.
Since we explained that very weak effect involving emissions and absorptions
of perturbative gravitons are consistent in the scheme of stochastic
quantisation, this may warrant at least formally the possibly unitarity in
the physics of the black hole evaporation. 

\section{Standard quantisation and first order stochastic equations}

We start now to explain how to reach concretely a second order Langevin equation for
quantum gravity.
We first explain the basic features of the first order stochastic quantisation,
the goal being the justification of formula~\eqref{gra}.

Let us consider a scalar field $\Ph(X)$ in $d$-dimensional Euclidean space
with coordinates $X^\mu$. 
Stochastic quantisation involves an additional ``stochastic time'' variable
$\tau$, in the sense that one promotes the field $\Ph(X)$ depending on $X$
to $\Ph(X,\tau)$ depending on both $X$ and a stochastic time $\tau$, and is
aimed at defining the $\tau$-dependent correlations functions
$\bra\Ph(X_1,\tau_1)\cdots\Ph(X_n,\tau_n)\ket$.
If the limit $\tau\to\infty$ exists, it must equate the Euclidean correlator
of the $d$-dimensional Euclidean QFT obtained by the standard formalism,
either canonical quantisation or the path integral, according to
\bq\label{ref0}
\lim_{\tau\to\infty}\bra\Ph(X_1,\tau_1)\cdots\Ph(X_n,\tau_n)\ket
\big|_{\tau_1=\cdots=\tau_n=\tau}\equiv\langle\Ph(X_1)\cdots\Ph(X_n)\rangle.
\eq
In zero dimensions, $\Ph(X)$ reduces to a point $x$ and $\Ph(X,\tau)$ is
simply a real function $x(\tau)$.
So, in zero dimensions, the concept of stochastic quantisation can be
handled with standard knowledge in analysis and distribution theory, and 
one gets the Brownian motion justification of the Boltzmann distribution.
The extension to QFT is a limit where the number of points becomes infinite,
but the general features of the zero dimensional case must remain the same.
Stochastic quantisation extends at the quantum level the classical notion of
a flow equation.

The quantum field theory proposal of Parisi and Wu \cite{parisi} is that the
stochastic time dependence of fields is through a (first order) Langevin
equation, where the drift force is the Euclidean equation of motion.
Giving a physical interpretation to the stochastic time $\tau$ is a 
step forward.
In fact, Langevin built its basic equation as a second order equation, for non
relativistic systems made of a large numbers of particles, but he reduced it
to a first order equation by neglecting the acceleration, which is often a
negligible term when one approaches the the thermal equilibrium.

Given a local action $S[\Ph]$ with the relevant properties, one may postulate 
the existence of a Fokker-Planck evolution kernel
$P^\FP(\Ph(X),\tau)$, such that 
\bq
\bra\Ph(X_1,\tau)\cdots\Ph(X_n,\tau)\ket\equiv
\mint{}{}[\d\Ph]_X\;\Ph(X_1)\cdots\Ph(X_n)\;P^\FP(\Ph(X),\tau).
\eq
Having a smooth quantum field theory means that $P^\FP(\Ph(X),\tau)$ satisfies
the parabolic equation
\bq\label{FP}
\mfrac{\pa}{\pa\tau}P^\FP[\Ph,\tau]=\mint{}{}\d X\mfrac{\del}{\del\Ph(X)}
\Big(\mfrac{\del S}{\del\Ph(X)}+\hbar
\mfrac\del{\del\Ph(X)}\Big)P^\FP(\Ph(X),\tau),
\eq
One must choose any given initial condition $P_0(\Ph(X))$ at $\tau=\tau_0$ for
$P^\FP$, but one can prove that the limit, if it exists, is independent of the
choice $P_0(\Ph,\tau_0)$, with an exponential damping in powers of
$\exp(-\tau)$ when $\tau\to\infty$.

The connection with Euclidean Feynman standard path integral is by proving
that the equilibrium distribution of the Fokker-Planck distribution is
\bq\label{FPE}
\lim_{\tau\to\infty}P^\FP[\Ph,\tau]=\exp-\frac 1 \hbar S[\Ph].
\eq
This is a stationary solution of Eq.~\eqref{FP}.
Gauge invariances can be taken into account in the stochastic framework.
However this necessitates a further degree of sophistication that involves 
the framework of gauge covariant TQFT's \cite{stoctqft}. 

It turns out that all relevant QFTs for elementary particles that are not
coupled to quantum gravity can be defined by such a first order
Fokker-Planck equation.
One gets therefore a probabilistic interpretation of all their Euclidean
correlation functions.

However, some Lagrangians, and in particular the Einstein Lagrangian, are
such that the limit \eqref{FPE} cannot be reached or does not make sense. 
Here we will suggest that this is not in contradiction with the possibility
that its correlators are well defined at finite values of the stochastic time,
with some possible oscillatory dependence when the stochastic evolution is 
generalised from first to second order.
The point of this paper it that this property has a relevant physical
interpretation.
 
A first order Langevin equation often implies a Fokker-Planck equation.
The Langevin framework is more general because it allows one to
define correlators at different values of the $\tau_i$.
It also a supersymmetric formulation in $d+1$ dimensions, whose
Hamiltonian interpretation illuminates very interestingly the Fokker-Planck
formulation of many well-behaved theories \cite{gozzi}.

As a matter of fact, Parisi and Wu proposed in their seminal paper
\cite{parisi} that the stochastic time dependence of the fields in the
correlation functions $\bra\Ph(X_1,\tau_1)\cdots\Ph(X_n,\tau_n)\ket$ is
determined from the equation\footnote{There is the possibility of inserting
a kernel in factor of $\mfrac{\del S[\Ph]}{\del\Ph(X,\tau)}$ to improve the
convergence, if any, of the stochastic process, without changing the
conclusion of the foregoing discussion.
A non-trivial kernel will appear in Eq.~\eqref{gra} and will be studied
in great details in \cite{bcw}.}
\bq\label{markov}
\mfrac{\pa\Ph(X,\tau)}{\pa\tau}=\mfrac{\del S[\Ph]}{\del\Ph(X,\tau)}
+\sqrt{\hbar}\,\eta(X,\tau),
\eq
where $\eta(X,\tau)$ is a noise whose correlation functions define the
quantisation, together with some initial condition of $\Ph$ at $\tau=\tau_0$,
say $\Ph(X,\tau_0)=\Ph_0(X)$.
In this simplest formulation leading to the Fokker-Planck equation~\eqref{FPE},
the noise is Gaussian white, which means
\bq
\bra\eta(X,\tau)\eta(X',\tau')\ket=2\del(\tau-\tau')\del(X-X').
\eq

The Langevin equation is thus very much like a Brownian motion equation where
$\frac{\del S[\Ph]}{\del\Ph(X,\tau)}$ is a drift force.
Eq.~\eqref{markov} shows that rotational invariance is generally not ensured
in $d+1$ dimensions.
Such an invariance can be however enforced in a few cases, e.g., for certain
Chern-Simons actions \cite{stoctqft}.
The stochastic time dimension is the square of the dimension of space
coordinates if the free part of the action is second order in the space
derivatives.

The prescription to get rid of the $\tau$ dependence and define the Euclidean
correlation functions
$\langle\Ph(X_1)\cdots\Ph(X_n)\rangle^{\mathrm{Euclidean}}$ of the
$d$-dimensional smooth quantum field theory is by solving eq.~\eqref{markov}
to express $\Ph$ as a function of $\tau$.
Then one computes the correlation functions of the $(d+1)$-dimensional theory
at equal $\tau$ by averaging over the noises and one takes the limit
$\tau\to\infty$, assuming for example that $\Ph=0$ at $\tau=0$.
The result is, generically,
\bq
\lim_{\tau\to\infty}\bra\Ph(X_1,\tau)\cdots\Ph(X_n,\tau)\ket=\mint{}{}
[d\Ph]_X\,\Ph(X_1)\cdots\Ph(X_n)\exp\big(\!-\sfrac{1}{\hbar}S[\Ph]\big)
\equiv\langle\Ph(X_1)\cdots\Ph(X_n)\rangle^{\mathrm{Euclidean}}.\nn
\eq

One can often prove the convergence of the stochastic process toward the same
limit for arbitrary initial conditions at $\tau=\tau_0$, whatever the value of
$\tau_0$ is.
The statistical Physics literature shows in many different ways that the
Langevin equation formulation gives the same result as the Fokker-Planck
formulation for $\lim_{\tau\to\infty}\bra f(\Ph_\tau)\ket$, but, as we
already said, the Langevin equation is more fundamental.
As will be explained in a separate publication, the Langevin equation
formulation allows one to investigate the interesting question of the
consequences of possible reparametrisations of the stochastic time.

In the Fokker-Planck formalism, it is rather easy to prove that the damping
of the Langevin/Fokker-Planck process toward the usual path integral formula
amounts to the possibility of normalising the Euclidean path integral, that is 
\footnote{The necessity of this condition is a general result of statistical
physics that can be proven in many ways. A simple proof is in \cite{huffel}}:
\bq\label{condc}
\mint{}{}[d\Ph]_X\,\exp\big(\!\!-\!\sfrac{1}{2\hbar}S[\Ph]\big)\,<\infty.
\eq
This condition is fulfilled for all renormalisable theories that satisfies
\eqref{condc} and
many quantum mechanical models with a discrete spectrum and a standard
normalisable vacuum.

But it is not fulfilled non-perturbatively in the case of $4$d gravity, since
in this case the field is $g_{\mu\nu}$ and $S[\Ph]$ amounts to the Euclidean
Hilbert-Einstein action that is not positive definite.
In this theory \eqref{condc} cannot hold because metric fluctuations can change
the sign of the scalar Euclidean space curvature. 
This is consistent with the fact that time cannot be globally well defined in
quantum gravity.

We now discuss how stochastic quantisation may improve our understanding of
quantum gravity.

\section{A possible way out for the stochastic quantisation of gravity}

For gravity, given some initials conditions to the Langevin equation, there is
generally no smooth limit for the stochastic process at $\tau\to\infty$
beyond perturbation theory.
We are lead to the following strategy. 

We observe that the stochastic evolution is well defined at finite values
of stochastic time. 
The metric correlators, generically denoted as
$\bra{\cal F}(g_{\mu\nu}(\tau,X))\ket$, describe some complicated $5$d physics
without possibility of defining a Lorentz time at finite values of $\tau$. 
But if the $4$d initial conditions corresponds to a semi-classical quantum
field theory configuration, the stochastic evolution of
$\bra{\cal F}(g_{\mu\nu}(\tau,X))\ket$ will admit a well-defined, smooth
$\tau\to\infty$ limit, which can described by a $4d$ semiclassical path
integral.

We thus we have the following possibility.
We can suppose that at some given finite value of $\tau$ there is a non-zero
probability that a given fluctuation transforms an arbitrary configuration in
a classical or semi-classical gravity configuration.
This transition can be seen as a reset of the boundary condition of the
stochastic process.
After this reset, the stochastic process becomes smooth by definition, since
the standard stochastic quantisation is compatible with the semi-classical
quantum field theory.

In order to trigger such a fluctuation, we need to use the exact Langevin
equation that contains a second order term, since it is only justified to
neglect it when one approaches smoothly the equilibrium limit for
$\tau\to\infty$. 
This is an advantage, since the energy carried by the stochastic acceleration
can enhance the radical changes in the metric to get a larger probability for 
the wanted transition to a semi-classical configuration.

Another remark is that the difficulties brought by the Wheeler-DeWitt equation
when one tries to quantise gravity in $4$d are of no relevance in the $5$d
formulation, because the gravitational Hamiltonian for the stochastic evolution
in $5$d does not vanish.
(As said earlier, there is no $5$d general covariance, and the Fokker-Planck
Hamiltonian is not vanishing, which implies that the $\tau$-evolution cannot
be frozen).
The stochastic time $\tau$ has thus a status different from that of the $4$
Euclidean coordinates $X^\mu$.
In \cite{bcw}, it will be shown in great details that the role of $\tau$ is
to foliate the $5$d space by Euclidean $4d$ leaves, each having its own $4$d
reparametrisation covariance.

When the convergence at $\tau\to\infty$ is ensured, one can in principle make
a Wick rotation that is an analytic continuation of all correlators in one of
the Euclidean coordinates $X^\mu$, in the limit $\tau\to\infty$.
Then all necessary Lorentzian correlators can be used to compute
perturbatively $S$-matrix elements, giving a particle interpretation to the
theory with well-defined scattering amplitudes and clocks.
The standard perturbation theory allows one to prove this property.

The question of the possibility of a Wick rotation at finite stochastic time
has not been extensively studied in the literature.
A recent paper has studied the question of establishing Hilbert space
positivity as a precise mathematical result at finite time \cite{jaffe}.
A simple argument is that it usually makes no sense to have a theory with two
distinct parameters of evolution to order phenomenon.

The equivalence proofs between standard path integral quantisation and
stochastic quantisation with a first order Langevin/Fokker-Planck amount to
show that there is a smooth damped relaxation of the stochastic process toward
the Euclidean correlation functions defined by the path integral with weight
$\exp(-\frac1\hbar S)$ (the analogue of a Boltzmann weight in statistical
physics).
It can only be done if the equilibrium process is itself well defined.

About the classical limit, when the effect of the noise is less stringent, the
solution is that the fields concentrate around the classical solutions, until
they reach the classical limit.
In fact, we call $\tau$-dependent classical states solutions to the Langevin
equations when the noise is neglected all along the $\tau$ evolution.
This operation can be heuristically understood by taking the limit $\hbar\to0$
in the Langevin equation for certain fields.
In fact, stochastic coherent states can be also defined as solutions that
depend on the noise and are as near as possible to stochastic classical
fields when one computes correlators in a natural generalisation of the
Schr\"odinger picture.

We pointed that, as in the theory of earthquakes, considering the effect of
the stochastic acceleration can be a useful feature with relevant consequences. 
Thus, we now come to the generalisation of a first order stochastic evolution
into a second order one.

\section{Stochastic quantisation with second order Langevin equations}
We now generalise the first order equation \eqref{markov} into
\bq\label{markov2}
\Del T^2 \mfrac{\pa^2\Ph(X,\tau)}{\pa^2\tau}
+2b\,\mfrac{\pa\Ph(X,\tau)}{\pa\tau}
=\mfrac{\del S[\Ph]}{\del\Ph(X,\tau)}+\sqrt\hb\,\eta_\Ph(X,\tau).
\eq
In other words, we will analyse the new physical features of the dynamics of
stochastic quantisation after $\pa_\tau\equiv\pa/\pa\tau$ is replaced by
$\Del T^2\,\pa_\tau^2+2b\,\pa_\tau$.
Here $b$ is a dimensionless positive real number, $a$ is a positive real
number with the same dimension as $\sqrt\tau$ and can be perhaps used as a
physical ultra-violet cutoff, $ \sim 1/\Del T$, for certain theories.
For $\Del T\to0$, the standard stochastic quantisation occurs, and it is a
consistent formulation provided we have a renormalisable QFT with an
appropriate potential.
If $b=0$, we have oscillatory solutions that are unlikely to becomes
stationary for $\tau\to\infty$.
So we must always have a friction term, proportionally to $b$.
A deeper understanding of the dissipation that occurs for $b\neq0$ is of
great interest.

For $\Del T\neq0$, as discussed in \cite{cosmob}, the standard first order
equation is effectively recovered when the action of the operator
$\Del T^2\pa^2_\tau$ is much smaller than the action of $2b\,\pa_\tau$, at
least perturbatively, if one has a renormalisable theory for $S$.
The combination of the effect of the noise and of the acceleration term
proportional to $\Del T^2$ can lead the system to nontrivial relaxations
towards a possible equilibrium, like oscillating ones that cannot be
predicted by a genuine first order equation.
This justifies heuristically the intuition that $1/a$ is a physical UV cutoff,
by comparing the time scale of stochastic time oscillations with those time
scales that occur in standard quantisation when $\tau\to\infty$.
In fact, beyond all heuristic justifications as those of \cite{cosmob}, the
original version of the Langevin equation was of second order.
Langevin approximated it as a a first order equation, only for the purpose of
studying the thermal equilibrium approach.

In the example of \cite{cosmob}, we considered a $6$-dimensional scalar
(one real and one complex field) renormalisable theory with interacting terms
$g\phi^3+e\phi\bar\psi\psi$.
(This was to avoid the more confusing case of fields with genuine
self-interactions.)
In this theory the real field $\phi$ can be be treated as a background
coherent state $\phi_\cl$ and $\psi$ is a complex quantum field that couples
with the background $\phi_\cl$ by its a current $e\bar\psi\psi$, similar to a
background electromagnetic field coupling to an electron current, or a
classical background gravitational field coupling to a microscopic quantum
black hole pair.

Eq.~\eqref{markov2} is translation invariant.
Since it contains the friction term proportional to $b$, the stochastic
process fundamentally involves a dissipation of the stochastic time energy.
For the cases where there is dumping toward the equilibrium at infinite
stochastic time, this stochastic energy will reach some constant.
For $a\neq0$, one has oscillations on the way, and the entropy increase
follows a different pattern.
The physics at finite stochastic time is different for $\Del T=0$ and
for $\Del T\neq0$.
Before the possible equilibrium, there are stochastic time slices during
which the term $\Del T^2\;\!d^2/dt^2$ dominates the term $b\;\!d/dt$.

Non-trivial physics occurs then with an approximate conservation of the
energy, since one can neglect the friction term for a while.
For these intervals of time, one can estimate which phenomena must balance
the $\tau$-energy variations in the $\tau$ oscillating background $\psi_\cl$.
The latter is computed by solving the Langevin equation for $\eta_\phi=0$,
and the quantum effects must be considered as those around this background,
which one understands how to compute.
Our scenario is that the quantum excitations for the field $\psi$ during this
$\tau$-dependent process are analogous to pair creations of the Schwinger
type.

The pair creations or absorptions that occur all along the $\tau$-evolution are
a back reaction that compensates the effects of the vacuum $\tau$-oscillations.
The whole process is driven by the effect of the noise and its relation to the
field by a second order differential equation in $\tau$.
The stronger $\phi_\cl$ is, the stronger is the counter effects of pair
annihilation and creation for the field $\psi$.
In principle, this can be perturbatively computed by solving the stochastic
equation including the effect of the noise $\eta_\psi$ in the background field
$\phi_\cl$.

When one relaxes to an equilibrium, if any, the damping of the periodic growth
and contractions of the energy of $\phi_\cl$ stops, as well as the creations
and absorptions of pairs as they given by the Langevin equation, whose
consequences can be described by a supersymmetric theory in the bulk. 
We will be more explicit in what is the supersymmetric theory in the bulk in
Appendix~B using a concrete and very simple example.
For gravity, we actually claim the equilibrium can be reached in a
discontinuous way, under the form of a phase transition, because the field
properties mix with the structure of the space. 
A comparison with what can happen with spin glasses was discussed in the
Introduction.

In Appendix~A, we shall justify some of these speculations by a careful
analysis of a zero dimensional case, where the fields $\Ph$ and $\Ps$ are
replaced by two real numbers $x$ and $y$, as an extreme dimensional reduction,
to explain the methodology and the way to handle the initial data.
It was argued in \cite{cosmob} that this early dynamics might even keep away
the field excitations from dangerous part of the potential for a while.

There is a period where the damping is very slow as compared to the rapidity of
the oscillations.
During this period, one may approximate
\bq
\Del T^2\mfrac{\pa^2\Ph(X,\tau)}{\pa^2\tau}+2b{\pa\Ph(X,\tau)}{\pa\tau}
\sim\Del T^2 \mfrac{\pa^2\Ph(X,\tau)}{\pa^2\tau}
\eq
The dashed line in the heuristic description by Figure~1 represents this
regime.

\begin{figure}\label{FIG}
\begin{center}
{\includegraphics[width=9cm]{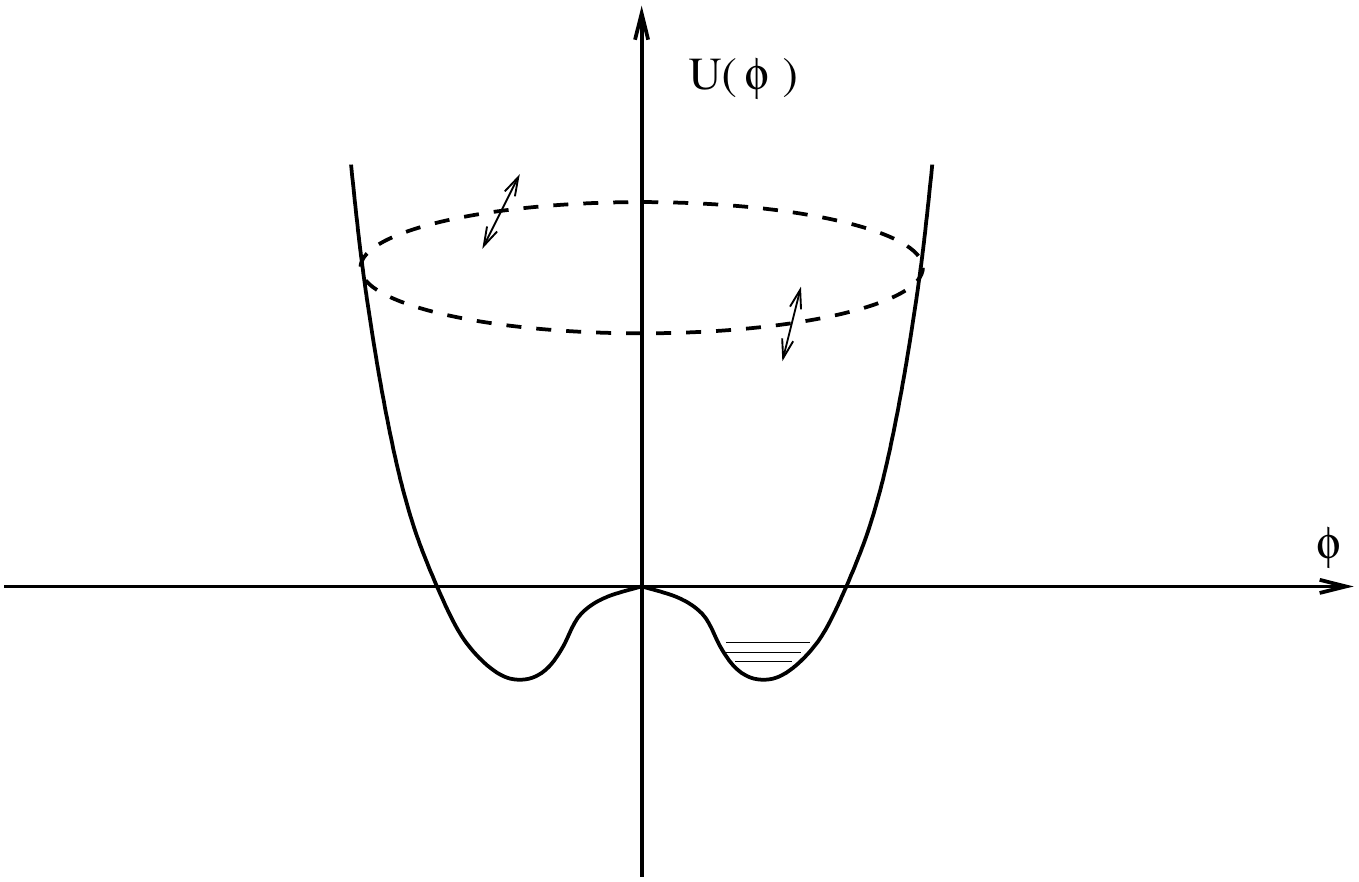}}
\quote{\small Figure~1: The dashed line represents one among the oscillating
stochastic time dependent ``vacua'' (OSTDV) $\phi_\cl$, and the double arrows
represent symbolically the pair creations and annihilations that occur to
periodically compensate for the approximate conservation of the total energy
when $\tau$ varies, as explained in the text.
So the ``inertial force'' due to the second order terms keeps $\phi_\cl$ 
away from the absolute minima of the potential for a while; the process
is reminiscent of a turbulent phenomenon.
Because of the friction terms in the second order Langevin equation, the
angular velocity decreases as $\tau\to\infty$ and $\phi_\cl$ will fall into
one of the absolute minima of the potential.
Then, either the ordinary quantisation or the classical behaviour will prevail
with the damping of the $\tau$ dependence.
For gravity, the equilibrium is a diluted Universe, filled with scattered
classical primordial black holes, the smallest of which decaying then quickly
to matter.}
\end{center}
\end{figure}

\section{Introducing the second order Langevin equation for the gravity
quantisation}

For gravity, one must promote the space metric $g_{\mu\nu}(X)$ to
$g_{\mu\nu}(X,\tau)$.
One needs to introduce of additional components $g_{\mu\tau}(X,\tau)$ and
$g_{\tau\tau}(X,\tau)$ following the general method for defining the
stochastic quantisation of a theory with a gauge invariance.
Thanks to these extra fields, one can construct a covariant derivative
$\nabla_\tau=\nabla/\partial\tau$, and the gravity stochastic equation can
be written using the method of equivariant cohomology.
(For a $2$d example and the YM$_4$ case where such additional fields are
introduced, see the last reference in \cite{stoctqft}.)
The second order gravity Langevin equation must be
\begin{align}\label{gra}
\Del T\nabla_\tau^2 g_{\mu\nu}(X,\tau,)+&b\nabla_\tau g_{\mu\nu}(X,\tau)
=G_{\mu\nu\mu'\nu'}(R^{\mu'\nu'}(X,\tau)-\mfrac12g^{\mu'\nu'}(X,\tau)R
\nonumber \\
&+\kappa\,g^{\mu'\nu'}(X,\tau)-8\pi G\,T^{\mu'\nu'})
+\sqrt \hbar\,\eta_{\mu\nu}(X,\tau),
\end{align}
 $\eta_{\mu\nu}$ is the noise for the Euclidean $g_{\mu\nu}$, and one
must have a (non-white) Gaussian distribution 
\bq
\bra\eta_{\mu\nu}(x,\tau)\eta_{\mu'\nu'}(x',\tau')\ket=
G^{-1}_{\mu\nu\mu'\nu'}\,\del^4(x-x')\del(\tau-\tau'),
\eq
where the kernel $G_{\mu\nu\mu'\nu'}(x,\tau)$ is a metric on the space of
metrics $g_{\mu\nu}(x,\tau)$ and is a quadratic functional of $g_{\mu\nu}$
for ensuring the reparametrisation invariance of \eqref{gra} for each value
of $\tau$.

Eq.~\eqref{gra} is a complicated non-linear differential equation.
It will be analysed in great details in \cite{bcw}, where most importantly,
the role and the exact definition of $G_{\mu\nu\mu'\nu'}$ will be thoroughly
examined.
The construction and geometrical identification of the covariant stochastic
speed and acceleration operators $\nabla_\tau$ and $\nabla_\tau^2$ is a
non-trivial task and will be also explained in \cite{bcw}.

If one solves Eq.~\eqref{gra} and eliminates the noise in the standard way as
prescribed by stochastic quantisation method, it produces correlation functions
at a finite stochastic time $\tau$ that depend on the initial data
$g_{\mu\nu}(X,\tau=\tau_0)$.
Therefore the dimensionful parameters that may occur in the solutions of
the stochastically quantised quantum gravity are not only $a$, $b$ and
$\hbar$, $\kappa$ and the Newton constant $G$, but also the initial condition
that describes the initial Euclidean geometry of the universe at $\tau=\tau_0$.

The value of the stochastic time, measured in units of $a$, when the Universe
exits from inflation depends on the initial data.
After the exit from inflation, which was induced by a random fluctuation
according to our scenario, gravity becomes by definition purely semi-classical.
One has essentially a standard classical evolution of the gravity from a given
geometry, modulo some negligible emissions of perturbative gravitons, which
must fit the data of today's observations.

The physical events, when gravity was behaving in quantum and non-perturbative
way, were ordered by the evolution in stochastic time in a $5$d framework.
It is only when the stochastic time has reached the value when gravity can be
treated effectively as a semi-classical theory everywhere can the correlators
in the final $4$d leave be computed by the standard path integrals with the
semi-classical gravity weight $\exp(-\frac1\hbar S)$.
All Physics can then computed by its value at the boundary $\tau\sim\infty$
and the $\Del T$ dependence is fully negligible.
Modulo a very short period where one needs inflaton fields, gravity can then
be treated for the whole future semi-classically while other interactions are
still treated fully quantumly.
It is only after this possible mathematically emergence of the Lorentz time,
by the possibility of a Wick rotation of of the Euclidean correlators in
stochastic equilibrium, that causality can be introduced in the usual sense,
giving the property that an observer can only detect phenomena having
occurred in its future light cone.

Before the transition between primordial cosmology and the post-inflation
semi-classical gravity regime, the impossibility of defining the Euclidean
path integral of gravity is due to the fact that the Einstein action is not
suited to define a normalised equilibrium description for $\hbar\neq0$.
In the language of statistical physics, it translates to saying that we
cannot compute by a Boltzmann average the correlators of a system that has
no thermalisation.

\section{Conclusion and physical outputs}
There is interesting physics for the correlators of a stochastically quantised
Euclidean field theory at finite stochastic time when stochastic quantisation
$\tau$ is of second order.
It is ignored because we cannot build clocks that measure the evolution in
function of the stochastic time.

We have done some elementary computations to illustrate the non-trivial
oscillations undergone by correlators when the stochastic process converges
smoothly.

In some cases, it may happen that no smooth limit $\tau\to\infty$ exists,
independent of the initial conditions.
This is the case for gravity.
Our proposal, with non-trivial physical consequences, is that instabilities
can occur on the way, especially when stochastic quantisation $\tau$ is of
second order, so that the phase of the system can change abruptly.

The Introduction of this article has detailed qualitatively some physical
predictions suggested by the definition of quantum gravity through stochastic
quantisation.
We will now make them a bit more detailed.

We suggest that stochastic quantisation opens a window for new speculations
on primordial cosmology.
It says that, given that the time scale $\Del T$ of stochastic oscillations 
is extremely small (e.g.,~$10^{-65}$s) compared to the scale of the length
of the exit of inflation $\sim10^{-32}$s, a fluctuation necessarily occur,
producing a transition from the quantum gravity phase to the classical phase.
The fundamental difference between the classical and the quantum phases is
the existence or not of the Lorentz time.

One must check the possibility of having a Lorentz time that emerges by
inspections of the analytical properties of the stochastic Euclidean
correlators that stochastic quantisation defines and allows one to compute.
This gives us two possible phases for gravity.
The classical phase, after the exit from inflation, is the standard model
coupled to classical gravity, where, effectively, all fields take their values
at infinite stochastic time, and all phenomena can be consistently ordered by
the Lorentz time, a property that can be proved theoretically (at least
rigorously in perturbation) theory, but that one must check experimentally,
instead of postulating.

The verification involves experiments done at effective infinite stochastic
time.
The accuracy of this verification is optimal in the big accelerators and in
astrophysics, and it goes down to $\sim 10^{-27}$s nowadays.
Atomic clocks provide an accuracy of $\sim 10^{-18}$s.
However, this accuracy is obviously totally out of scale as compared to the
smallness of the time scale for detecting quantum gravity
effects, as predicted by the Langevin equation.
Any given experiment is helpless, since it has to be made with an
instrumentation that is described by the physics of matter and fields that is
only relevant in our (semi)classical gravity phase.
This is unfortunate because the possible physical existence of the stochastic
time seems theoretically favoured, giving a microscopic justification of the
success of standard QFTs at the scale of all known experiences.

The other phase, where quantum gravity is involved, predicts that there is no
Lorentz time, which is in fact in agreement with the Wheeler-DeWitt
constraint.
Rather, there are oscillations in the stochastic time of Euclidean
gravitational correlators with no smooth relaxation, which defines the
physics of quantum gravity, and which we have pictured as exchanges of energy
between the $5$d classical solutions and the $5$d quantum states (whose $4$d
wave functions could be related to that of a $4$d instantons), as a
generalisation of the Schwinger effect.

So, we have suggested that there are oscillations in the stochastic time of
Euclidean correlators with no smooth relaxation, which define the physics of
quantum gravity.
The evolution in function of the stochastic time $\tau$ of this physics is
governed by the $4$d Hamiltonian ${\cal H}_4$ of the $5$d $(+----)$
supersymmetric theory
corresponding to the stochastic Langevin equation defined by Eq.~\eqref{gra}.
Using Legendre transform, ${\cal H}_4$ can be computed from the second order
Fokker-Planck supersymmetric action with metrics signature $(+----)$ of the
$4$d Euclidean gravity.
Fig.~1 sketches the oscillating exchanges of energy between the stochastic
time dependent coherent state obtained by solving the Langevin equation at
vanishing noise, which generalises into a $5$d background in the case of
gravity, with the annihilations or destructions of the $5$-dimensional states
around it, which are elements of the spectrum of ${\cal H}_4$.

This regime of oscillations can be abruptly changed with a transition from the
quantum to the classical phase provoked by a strong enough fluctuation,
acting as a reset of the initial conditions of the stochastic process.

The way we visualise the mechanism of stochastic quantisation of gravity in a
small enough volume implies that this transition results in the appearance of a
diluted Universe, with a rather small dimension at the beginning, and where the
Lorentz time has emerged (as a change in the analytical properties of
Euclidean correlation functions).
At the end of the inflation, it says that this small enough Universe is filled
with faraway scattered primordial black holes before it expands.
We may consider this as a physical prediction of the definition of Euclidean
quantum gravity through a second order Langevin equation.

The spatial distribution and mass distribution of theses primordial black
holes (meaning solutions of 4d classical Lorentz equations of motion) allow
a classical description of gravity almost everywhere.
It might be that this scheme gives gives a preferred initial condition for
any model based on inflation theory.
At this stage of the evolution of the Universe, it is by the definition of the
change in phase that the recently ``emerging'' Lorentz time allows us to
order phenomena by the standard QFT formalism with its own smooth causality.
The inflation theory gives a phenomenological description of the last moments
of the transition, as soon as the Lorentz time has emerged and can be used
to order phenomena.

After the transition, the mass distribution of these $4$d primordial black
holes opens the possibility of having elementary particles in between, under
the form of ordinary matter resulting from evaporation mechanisms of the
lightest primordial black holes.
In this $4$d description, one assumes that one never encounters the unsolved
question of $4$d point singularities. This assumption is justified by the
impossibility of concentrating enough energy at a given point to ignite the
quantum gravity regime.

Giving such justified initial conditions for the post inflation Universe
may provide a realistic description of the phase of the Universe as it is now,
using the Lorentz time as an effective parameter for the standard causal
evolution.
The inflation theory gives a phenomenological description of the last moments
of the transition, as soon as the Lorentz time has emerged and can be used
to order phenomena.

In fact, for these last moments of the transition, one may say that the
phenomenological smooth Lorentz time dependence is described by the choice
of this inflaton field.
The transition is definitely over when the cosmological constant has become
a fixed number and every manifestation of the former phase involving quantum
gravity has faded away.
Then, the inflaton field disappears from our description of physics, and all
fundamental fields are enough to describe physics by the standard QFT
formalism, where the stochastic time evolution has been thermalised, meaning
that effectively physics can be always computed at $\tau=\infty$, and the
Lorentz time can be used as an (almost) fundamental evolution parameter.
One may presumably find realistic and simple distributions of these predicted
primordial black holes with the present cosmological measurements as consistent
initial conditions for the various cosmological inflation models.

It is worth trying to give a suggestive name to the $5$d gravitational quantum
states that are created and absorbed in the $5$d space with signature
$(+- - - -)$.
It would be misleading to call them $5$d black holes; rather they are more like
particles in $5$d, with a $4$d particle shape.
We suggest to call them gravitational partons.
Before the phase transition, the metric has an oscillating evolution in
stochastic time.
It goes together with fast creations and annihilation of these $4$-dimensional
particles, which build the spectrum of ${\cal H}_4$.

Now, in a sort of reverse way, let us consider a gedanken experiment in the
$4$d Lorentz space.
Suppose we can build an accelerator to produce a collision at ultra short
distances of two massive point-like particles existing in our phase, and
observe their collision from afar.
We can for example consider the ultra-deep inelastic collision
$e+e\to\rm\ inclusive\ set\ of\ observed\ particles$.
When both particles are at an ultra short distance from each other, each one
``feels'' the very strong gravitational field created by the other particle.
This very short distance collision should be analysed in the context of quantum
gravity.
In the range of the interaction, the conditions are basically locally the same
as they are everywhere in primordial cosmology, and the theory that one uses
must the one we just discussed, with its $5$d metrics signature $(+----)$.

It is suggestive that, for the very short range collision, each particle sees
the other particle as a cloud of the above mentioned elements of the spectrum
of ${\cal H}_4$, and the interaction for the very short duration of the
collision amounts to interactions between these microscopic constituents,
governed by the $5$d theory, which we just called $5$d gravitational partons.
In this description of the ultra high energy collision of $4$d ``point-like"
particles, after the very short moment of the quantum gravity interactions,
the energy that was exchanged between the $5$d gravitational partons of both
particles will be redistributed among standard $4$d particle degrees of
freedom, which will be detected sufficiently far away from the zone of
interaction.
This view is a wild generalisation of the parton model in QCD.

This illustration of what might be the ultra short collisions of ``point-like"
particles might perhaps justify a posteriori the transition between the
primordial cosmology and the post-inflation period.

We can now rephrase the scenario.
At the end of the phase change that terminates the epoch of primordial
cosmology, namely the exit of the inflation, the states made of the $5$d
gravitational bound states are replaced by solutions of classical gravity
with the Lorentz signature $(+---)$.
When one observes them at a far enough distance, some of them behave as
ordinary $4$d elementary particles and others are organised as classical $4$d
black holes.
At the very beginning, it is most likely that we only have black holes, till
the Universe is large enough.
For all of them, effectively, the stochastic time can be estimated as blocked
at its infinite value.
Sufficiently far away from their singularities, we just have the
semi-classical gravity coupled to the rest of the standard interactions.
The evolution of the Universe can then be ordered in function of the Lorentz
time, with an effective field theory which is that of the inflation.
After the transition to its semi-classical phase, the microscopic quantum
gravity $5$d theory with the $(+----)$ signature is effectively reduced to a
$4$d theory with a $(+---)$ signature, modulo a Wick rotation from a $(----)$,
namely the standard semi-classical gravity.

After the exit from inflation, and for all realistic experiences, the standard
semi-classical gravity is helpless to describe the ``interior'' of the
elementary particles. 
This is compatible with the fact that the $4$d description of quantum gravity
cannot describe the physics at a very short distance such as the scale
$\Del T$.
In fact, our model says that the internal QFT ``inside'' a so-called ``point
particles'', e.g., an electron or a quark, is the $5$d quantum gravity theory.

The $5$-dimensional properties imply that all particles have a gravitational
structure and suppress the contradictions brought by the notion of being $4$d
point-like.
This also eliminates the question of trying to give a physical sense to the
gravitational singularities at a very short distance in $4$d with a Lorentz
metric: the gravity theory changes at extremely short distance in the $4$d
Lorentz space and becomes $5$d.
This theoretical extension might be useful to clarify aspects of the
information paradox.
Indeed in the $5$d formulation, no particle from the outer space will ever
penetrate in the usual sense the very short distance core of a black hole, at
the scale where the gravity becomes quantum without leading to contradiction
at long distances.
It might be the case that all particles must have at least a very tiny mass in
order to have a consistent QFT description in our $5$d formulation.

\bigskip

\bigskip\noindent {\bf Acknowledgment.}
Laurent Baulieu wishes to thank Giorgio Parisi, John Iliopoulos and
Benoit Dou\c cot for very encouraging discussions as well as the NCTS and its
Director Chong-Sun Chu in Hsinchu for its generous and warm hospitality.
SW is supported in part by grant No.~106-2115-M-007-005-MY2 from MOST (Taiwan).

\vskip1cm

\appendix
\noindent{\bf\Large Appendix}
\section{Second order Langevin equation in a perturbative zero-dimensional
example}
Consider the zero dimensional case of two real variables $x$ and $y$ with
an action
\bq
S(x,y)=\mfrac12M^2x^2+\mfrac12m^2y^2+\mfrac{\lam}{2}xy^2.
\eq
The stochastic quantisation of $S(x,y)$ introduces a bulk time $\tau$ and
promotes $x,y$ to $x(\tau),y(\tau)$, and one builds eventually a quantum
mechanical path integral with a supersymmetric Lagrangian
${\cal L}_\stoc(x,y,\psi_x,\psi_y)$, to be determined later on, such that
for any observable $f(x,y)$, one has
\begin{align}\label{mention}
&\qquad\int \d x\d y\;f(x,y)\,\exp\big(\!\!-\!\sfrac{1}{\hb}S(x,y)\big) \nn\\
&=\lim_{T\to\infty}\int[\d x]_\tau[\d y]_\tau[\d\psi_x]_\tau[\d\psi_y]_\tau
[\d\bar\psi_x]_\tau[\d\bar\psi_y]_\tau\;f(x(\tau),y(\tau))\,
\exp\Big(\!\!-\!\sfrac1\hb\mint0T\!\d\tau\,
{\cal L}_\stoc(x,y,\psi_x,\bar\psi_x,\psi_y,\bar\psi_y)\Big).
\end{align}
Second order stochastic quantisation means in fact that, in correspondence
with such a supersymmetric representation, there is an underlying Langevin
equation
\begin{align}\label{stof}
\Big(a^2\mfrac{\d^2}{\d\tau^2}+2b\,\mfrac{\d}{\d\tau}\Big)\,x(\tau)
+M^2x(\tau)+\mfrac{\lam}{2}\,y^2(\tau)&=\sqrt\hb\,\eta_x(\tau), \nn\\
\Big(a^2\mfrac{\d^2}{\d\tau^2}+2b\,\mfrac{\d}{\d\tau}\Big)\,y(\tau)
+m^2y(\tau)+\lam\,x(\tau)y(\tau)&=\sqrt\hb\,\eta_y(\tau)
\end{align}
that defines the $\tau$ evolution.
Here $\eta_x$ and $\eta_y$ are taken as Gaussian noises, i.e.,
\bq\label{eta^2}
\bra\eta_x(\tau)\ket=\bra\eta_y(\tau)\ket=0,\qquad
\bra\eta_x(\tau)\eta_x(\tau')\ket=\bra\eta_y(\tau)\eta_y(\tau')\ket
=2b\del(\tau-\tau').
\eq
All other Gaussian correlators for mean values of higher order products of
the $\eta$'s follow by averaging with the Gaussian distribution
$\int[\d\eta]_\tau\exp(-\frac{1}{4b}\int\d\tau\,\eta^2(\tau))$.

The case $a=0$ and $b=\frac12$ is for the standard first order stochastic
quantisation of \cite{parisi}.

In what follows, both $a\neq0$ and $b\neq0$.
For a more general potential $S=\frac12M^2y^2+\frac12m^2x^2+\lam V(x,y)$,
we must replace $\frac{\lam}{2}y^2$ and $\lam xy$ in Eq.~\eqref{stof} by
$\lam V_x$ and $\lam V_y$, respectively.
Of course, not all choices of the ``pre-potential'' $S$ produce proper
convergence of solutions at infinite $\tau$.
For $a\neq0$, we need to complete each Langevin equation with two boundary
conditions, instead of one condition in the first order case.
Here we assume that at two values of the time, $\tau=\tau_1$ and $\tau_2$,
the coordinates $x$ and $y$ take the values $x(\tau_i)=x_i$ and
$y(\tau_i)=y_i$ for $i=1,2$.
For regular theories, the infinite time limit is expected to be independent
of the choice of such conditions.

Our aim is to consider that one of the field, here $x(\tau)$, is a
``coherent state'', which is, in the Schr\"odinger sense, a state that
minimises maximally some quantum fluctuations.
So $x(\tau)$ is a state that is as close as possible to a solution where
one neglects everywhere $\eta_x$.
This situation has been advocated to in \cite{cosmob}, to define the
primordial cosmology.

Moreover, since we understand intuitively that at finite stochastic time
$y(\tau)$ undergoes quantum effects around the ``strong'' coherent state
$x(\tau)$, we introduce an arbitrary given fixed position $x_\cl$ for $x$,
which can be interpreted as the classical equilibrium value of
$\bra x(\tau)\ket$ when $\tau\to\infty$.
We thus reduce the above coupled Langevin equations, to the case where
effectively $\eta_x=0$, that is,
\begin{align}\label{L2}
a^2\ddot x+2b\,\dot x+M^2(x-x_\cl)+\mfrac{\lam}{2}\,y^2&=0, \nn\\
a^2\ddot y+2b\,\dot y+m^2y+\lam\,xy&=\sqrt\hb\,\eta_ y.
\end{align}

Solving perturbatively Eqs.~\eqref{L2} as a Taylor expansion on $\lam$
is of course possible.
Using Green's function techniques gives a hint of the physics at finite $\tau$.
In fact, some aspects of the loop expansion that will occur for $a\neq0$ are
better viewed by the redefinition of fields and coupling constant,
\bq
\hbar\lam\to\lam,\qquad\mfrac{x}{\sqrt\hb}\to x,\qquad\mfrac{y}{\sqrt\hb}\to y.
\eq
After such redefinitions, the Planck constant $\hb$ disappears from the
Langevin equations, which become
\bq\label{L22}
D^M_\tau(x-x_\cl)+\mfrac{\lam}{2}y^2=0,\qquad D^m_\tau y+\lam xy=\eta_y,
\eq
where $D^M_\tau\equiv a^2\,\pa_\tau^2+2b\,\pa_\tau+M^2$,
$D^m_\tau\equiv a^2\,\pa_\tau^2+2b\,\pa_\tau+m^2$, and the perturbative loop
expansion of the correlators can be expanded in powers of the rescaled
coupling constant $\lam\hbar\to\lam$.

\subsection{Perturbative expansion and finite $\tau$ QFT behaviour}
The perturbative expansion in $\lambda$ for the $\tau$ evolution of $x$ and
$y$ can be represented by Feynman diagrams with insertions of the noises
analogous to those in \cite{parisi}, except that its propagators $G_M$ and
$G_m$ have a double pole structure instead of being of parabolic type in
\cite{parisi}.
There is a forward propagation of modes with positive and negative energy
in the $\tau$ evolution, and one has insertions of the field $x(\tau)$ in
addition to the insertions of $\eta_x$ on the propagators of the $y(\tau)$.
The Feynman diagrams that one can draw to describe perturbation theory
involve closed loops.

In a Fourier transformation over $\tau$, using the conjugate variable $E$,
we must define particles of type $y$ and antiparticles of type $\bar y$,
with creation and annihilation operators acting on a Fock space, basically
because we have solutions of positive and negative energy in a symmetric way.

Because $a\neq0$, the $\tau$ dependence implies a relativistic quantum field
theory framework rather than a non-relativistic framework as in the case $a=0$.
The closed loops can be interpreted as forward stochastic time propagations of
particles of type $y$ and antiparticles of type $\bar y$.

The closed loops are finite integrals over a one dimensional momentum space,
with neither infra-red nor ultra-violet divergences, since $m\ne0$ and $M\ne0$.
They occur in the perturbative expansion of $\bra x^p(\tau)y^q(\tau)\ket$ and
can be interpreted as creations of a virtual pairs created by the vertex
$\lam xy\bar y$ of particle and antiparticle $y$ and $\bar y$ at a given value
of the stochastic time, each one propagating forwardly in $\tau$, until they
annihilate at a further stochastic time, with possible interactions with the
``classical field'' $x(\tau)$.

This suggests that a double Fock space must be constructed.
It is made of all possible states that can occur for the $\tau$ evolution,
one for all possible ``background vacua'' of $x$, representing the stochastic
oscillations of $x$ and determined by solving the second order Langevin
equation at $\eta_x=0$, and the other one for the ordinary quanta of the
field $y$ emitted around this background.
This description will become clearer by studying the Fokker-Planck Lagrangian
${\cal L}_\stoc$ and Hamiltonian associated to the second order Langevin
equation: ${\cal L}_\stoc$ contains a higher order derivatives in the
stochastic time, and implies a doubled phase space in the (stochastic time)
Hamiltonian formalism.

Here we use the approximation that the field $x(\tau)$ is a coherent state,
made up of its elementary quanta, in a way that minimises the uncertainty
principle.

The elementary quantum processes that build the perturbation theory are the
possible decay, annihilation and diffusion reactions
\bq
x_\cl\to y+\bar y,\qquad y+\bar y\to x_\cl,\qquad y+x_\cl\to y+x_\cl,
\qquad\bar y+x_\cl\to\bar y+x_\cl
\eq
whose strength is proportionally to $\lambda$.
The $\tau$ translation symmetry of the Langevin equation implies that at each
vertex there is a conservation of the $\tau$ energy, so that the $\tau$
evolution of the field $x_\cl$ can be accompanied by real decay and real
annihilations of pairs of $y$ and $\bar y$ quanta, namely by a Schwinger type
mechanism.

Because we have a friction term proportional to $b$, the phenomena that occur
during the $\tau$ evolution will disappear in the limit $\tau\to\infty$, if
it exists.
This is the case in our example.

Once $x(t)$ and $y(t)$ have been diagrammatically expressed at a given order
of perturbation, one can compute at the same order of perturbation theory
$x^p(\tau)y^q(\tau)$, and then averaging, one can obtain
$\bra x^p(\tau)y^q(\tau)\ket$ by using the fact that $\eta$ has a Gaussian
distribution.

Thus, there is a perturbation expansion involving interactions and propagators,
with closed loops, which determines for every value of $\tau$ the expectation
values $\bra x^p(\tau)y^q(\tau)\ket$ as a formal series in $\lam$, and one can
compute it at any given finite order in $\lam$.
The final result is expressed as
\begin{align}\label{gaussian}
&\lim_{T\to\infty}\bra x^p(\tau)y^q(\tau)\ket  \nn \\
&=\mint{}{}\d x\d y\,x^py^q\,\del(x-x_\cl+o(\lam))\exp\big(\!-\sfrac12
(M^2(x-x_\cl)^2+\sfrac12m^2y^2+\sfrac12\lam xy^2)(1+o(\lam^2))\big) \nn \\
&=Z_x^px^p_\cl\mint{}{}\d y\,y^q\,
\exp\big(\!-\sfrac12Z_m^2m^2y^2+\sfrac12Z_{\lam}\lam x_\cl\,y^2\big).
\end{align}
The $Z$ factors are finite renormalisation factors with a Taylor expansion,
$Z=1+\lam Z^1+\lam^2Z^2+\cdots$, where all the finite coefficients $Z^n$ can
be in principle computed in the perturbation theory in $\lam$.

The last formula tells us that Langevin equations gives us a very complicated
way to define a standard Gaussian integral in zero dimensions, with a well
defined theory that computes the corrections in $\lam$ at any given finite
order of perturbation theory!
The way the limit is reached is an exponential damping, with a regime of very
fast oscillations, as sketched in the figure \eqref{FIG}.

We will check this explicitly for the one point and two point functions
$\bra x(\tau)\ket$ and $\bra y(\tau)y(\tau')\ket$.

\subsection {0-loop and 1-loop computations of two point functions}\label{01}
The Green's function $G^M(\tau)$ of the operator
$D_\tau^M=a^2\pa_\tau^2+2b\pa_\tau+M^2$ satisfying $D_\tau^MG^M=\del(\tau)$
can be computed by a Laplace or Fourier transform.
Suppose $aM>b>0$.
It is
\bq
G^M(\tau)=\theta(\tau)
\mfrac{\exp(-E_+^M\tau)-\exp(-E_-^M\tau)}{a^2(E_-^M-E_+^M)}
=\mfrac{i\,\theta(\tau)}{2\sqrt{a^2M^2-b^2}}
\big(\!\exp(-E_+^M\tau)-\exp(-E_-^M\tau)\big),
\eq
where
\bq
E_\pm^M=\sfrac1{a^2}\big(b\pm i\sqrt{a^2M^2-b^2}\,\big)
\eq
satisfy $a^2E^2+2bE+M^2=(E+E_+^M)(E+E_-^M)$.

When $a\neq 0$, the free propagator for the $\tau$ evolution has still an
exponential damping factor, with a characteristic time that is proportional
to $b^{-1}$ (when $\tau$ is counted in units of $a^2$), but there is a new
phenomenon, which are $\tau$-oscillations that can be of a very high frequency
(in units of $a$) if the mass $M$ is large enough.

This is the situation that was suggested generically in \cite{cosmob}.
In the case of QFTs, one should replace $M^2$ by $M^2+\vec k^2$, where
$\vec k$ stands for the momentum of the particle.
Care must be given to the possible UV divergency when $\vec k^2$ becomes
very large.

If $am>b$, we have $G^m(\tau)$ and $E_\pm^m$ similarly for the operator
$D_\tau^m=a^2\pa_\tau^2+2b\pa_\tau+m^2$.

We consider the coupled Langevin equations~\eqref{L22}, with only one noise
for $y(\tau)$.
We solve Eq.~\eqref{L22} perturbatively.
Suppose
\bq
x(\tau)=x_0(\tau)+\lam x_1(\tau)+o(\lam^2),\qquad
y(\tau)=y_0(\tau)+\lam y_1(\tau)+o(\lam^2).
\eq
satisfy \eqref{L22}.
Then the $0$th order terms in $\lam$ satisfy
\bq\label{0th}
D_\tau^M(x_0-x_\cl)=0,\qquad D_\tau^my_0=\eta
\eq
whereas the first order terms in $\lam$ satisfy
\bq\label{1st}
D_\tau^Mx_1+\sfrac12 y_0^2=0,\qquad D_\tau^my_1+x_0y_0=0.
\eq
The solution to the first (homogeneous) equation for $x_0(\tau)$ in
\eqref{0th} is
\bq x_0(\tau)=x_\cl+c_+^M\exp(-E_+^M\tau)+c_-^M\exp(-E_-^M\tau)
=x_\cl+o(e^{-\tau}), \eq
where $c_\pm^M$ are constants that are determined by the chosen values of
$x_0$ at some $\tau_1$ and $\tau_2$, and $o(e^{-\tau})$ stands for any term
that is dominated by $e^{-\epsilon\tau}$ for some $\epsilon>0$ (including
the oscillations) as $\tau\to\infty$.
Thus
\bq\label{<x0>}
\bra x_0(\tau)\ket=x_\cl+o(e^{-\tau}).
\eq

Indeed, when the stochastic process converges, the limit at infinite
stochastic time of correlators of the fields are independent of the chosen
values of the fields at $\tau_1$ and $\tau_2$ as well.
Since $E^M_\pm$ have a positive real part, the damping in the dependence of
boundary conditions is exponentially fast in $\tau$, times some oscillations.

On the other hand, the equation in \eqref{0th} for $y_0(\tau)$ is
inhomogeneous and we have
\bq
y_0(\tau)=(G^m*\eta)(\tau)+c_+^m\exp(-E_+^m\tau)+c_-^m\exp(-E_-^m\tau)
=(G^m*\eta)(\tau)+o(e^{-\tau})
\eq
for some constants $c_\pm^m$, where $*$ stands for the convolution.
Similarly, $c_\pm^m$ are related to the boundary values of $y_0(\tau)$ at
some $\tau_1$ and $\tau_2$, and their choice does not affect the correlators
at infinite stochastic time.
If $0<\tau_1\le\tau_2$, we have
\begin{align}
\bra y_0(\tau_1)y_0(\tau_2)\ket&=\mint0{\tau_1}\d\tau'_1\mint0{\tau_2}\d\tau'_2
\;G^m(\tau_1-\tau'_1)G^m(\tau_2-\tau'_2)\,\bra\eta(\tau'_1)\eta(\tau'_2)\ket
+o(e^{-\tau_1})    \nn \\
&=2b\mint0{\tau_1}\d\tau'\,G^m(\tau_1-\tau')G^m(\tau_2-\tau')+o(e^{-\tau_1})
    \nn \\
&=\mfrac{b}{a^4(E_+^m+E_-^m)(E_-^m-E_+^m)}
\Big(\mfrac{\exp(-E_+^m(\tau_2-\tau_1))}{E_+^m}
-\mfrac{\exp(-E_-^m(\tau_2-\tau_1))}{E_-^m}\Big)+o(e^{-\tau_1}).
\end{align}
In particular, taking $\tau_2=\tau_1=\tau$, we have
\bq\label{<y0^2>}
\bra y_0(\tau)^2\ket=\mfrac{b}{a^4(E_+^m+E_-^m)E_-^mE_+^m}+o(e^{-\tau})
=\mfrac1{2m^2}+o(e^{-\tau}).
\eq

For the next order, we solve $x_1(\tau)$ from Eq.~\eqref{1st} and obtain
\bq
x_1(\tau)=-\mfrac12\mint0\tau\d\tau'\,G^M(\tau-\tau')y_0(\tau')^2+o(e^{-\tau}).
\eq
Taking the expectation value, we have
\begin{align}\label{<x1>}
\bra x_1(\tau)\ket
&=-\mfrac1{4m^2}\mint0\tau\d\tau'\,G^M(\tau-\tau')+o(e^{-\tau}) \nn\\
&=-\mfrac1{4m^2}\cdot\mfrac1{a^2(E_-^M-E_+^M)}
\Big(\mfrac1{E_+^M}-\mfrac1{E_-^M}\Big)+o(e^{-\tau})  \nn\\
&=-\mfrac1{4m^2M^2}+o(e^{-\tau}).
\end{align}
Similarly, solving $y_1(\tau)$ from Eq.~\eqref{1st}, we obtain
\begin{align}
y_1(\tau)
&=-\mint0\tau\d\tau'\,G^m(\tau-\tau')x_0(\tau')y_0(\tau')+o(e^{-\tau}) \nn\\
&=-x_\cl\mint0\tau\d\tau'\,G^m(\tau-\tau')y_0(\tau')+o(e^{-\tau}).
\end{align}
Therefore, taking the expectation value, we have
\begin{align}\label{<y0y1>}
&\quad\bra y_0(\tau)y_1(\tau)\ket   \nn\\
&=-x_\cl\mint0\tau\d\tau'\,G^m(\tau-\tau')\bra y_0(\tau)y_0(\tau')\ket \nn\\
&=\mfrac{bx_\cl}{a^4(E_+^m+E_-^m)(E_-^m-E_+^m)}
\mint0\tau\d\tau'\;G^m(\tau-\tau')
\Big(\mfrac{\exp(-E_+^m\tau')}{E_+^m}-\mfrac{\exp(-E_-^m\tau')}{E_-^m}\Big)
+o(e^{-\tau})    \nn\\
&=-\mfrac{bx_\cl}{2a^6(E_+^m+E_-^m)(E_+^mE_-^m)^2}+o(e^{-\tau}) \nn\\
&=-\mfrac{x_\cl}{4m^4}+o(e^{-\tau}).
\end{align}

Combining the above $0$th order and the $1$st order contributions to the
correlators, we have
\begin{align}\label{<x>}
\lim_{\tau\to\infty}\bra x(\tau)\ket&=\lim_{\tau\to\infty}\bra x_0(\tau)\ket
+\lam\lim_{\tau\to\infty}\bra x_1(\tau)\ket+o(\lam^2) \nn \\
&=x_\cl-\mfrac\lam{4m^2M^2}+o(\lam^2)
\end{align}
from Eqs.~\eqref{<x0>} and \eqref{<x1>}, and
\begin{align}\label{<y^2>}
\lim_{\tau\to\infty}\bra y(\tau)^2\ket&=
\lim_{\tau\to\infty}\bra y_0(\tau)^2\ket+2\lam
\lim_{\tau\to\infty}\bra y_0(\tau)y_1(\tau)\ket+o(\lam^2) \nn \\
&=\mfrac1{2m^2}+2\lam\cdot\Big(\!-\mfrac{x_\cl}{4m^4}\Big)+o(\lam^2) \nn \\
&=\mfrac1{2(m^2+\lam x_\cl)}+o(\lam^2)
\end{align}
from Eqs.~\eqref{<y0^2>} and \eqref{<y0y1>}.
In the limit, the $a$ and $b$ dependence is washed away, as well as that on
the initial conditions.

The shifts $x_\cl\to x'_\cl=x_\cl-\lam/4bm^2M^2$ and
$m^2\to m'\,{}^2=m^2+\lam x_\cl$ in Eqs.~\eqref{<x>} and \eqref{<y^2>} have
a simple explanation.
In Eq.~\eqref{L22}, substituting the expectation value $\bra y_0^2\ket$ in
\eqref{<y0^2>} for $y^2$ in the equation for $x$, we have
\bq D_\tau^M(x-x_\cl)+\mfrac\lam2\,\bra y_0^2\ket=D_\tau^M(x-x'_\cl) \eq
while substituting the expectation value $\bra x_0\ket$ in \eqref{<x0>} for
$x$ in the equation for $y$, we have
\bq D_\tau^my+\lam x\,\bra y_0\ket=D_\tau^{m'}y. \eq
Thus we have the same form of Eq.~\eqref{L22} but with the shifted constant
parameters $x'_\cl$ and $m'\,{}^2$.
This is consistent with Eq.~\eqref{gaussian} with the finite
renormalisation constants
\bq
Z_x=1-\mfrac\lam{4bm^2M^2x_\cl}+o(\lam^2),\quad
Z_m=1+\mfrac{\lam x_\cl}{2m^2}+o(\lam^2),\quad Z_\lam=1+o(\lam).
\eq

\subsection{Higher loops and diagrammatic expansions}
Recall the system \eqref{L22} of stochastic equations.
Iteratively, we have
\begin{align*}
x&=x_\cl-\mfrac\lam2 G^M*y^2=x_\cl-\mfrac\lam2 G^M*(G^m*\eta-\lam G^m*(xy))^2\\
&=x_\cl-\mfrac\lam2 G^M*(G^m*\eta)^2
+\lam^2x_\cl\,G^M*((G^m*\eta)(G^M*G^m*\eta))+o(\lam^3)
\end{align*}
and
\begin{align*}
y&=G^m*\eta-\lam G^m*(xy)=G^m*\eta-\lam G^m*\big(
\big(x_\cl-\mfrac\lam2 G^M*y^2\big)(G^m*\eta-\lam G^m*(xy))\big) \\
&=G^m*\eta-\lam x_\cl\,G^m*G^m*\eta+\lam^2x_\cl^2\,G^m*G^m*\eta
+\mfrac{\lam^2}2G^m*((G^M*(G^m*\eta)^2)(G^m*\eta))+o(\lam^3).
\end{align*}
Diagrammatically, the above expansions of $x$ and $y$ can be represented by

\begin{tikzpicture}[segment amplitude=2pt]
\draw(0,0)node{$x(\eta)\;\;\;\;=$};
\draw(1.5,0)circle(.1);
\draw(1.5,-1)node{\scalebox{.8}[.8]{$\lam^0$}};
\draw(2.25,0) node{$+$};
\draw[snake=snake,segment length=4pt](3,0)--(4.25,0);
\draw(4.25,0)--(5.5,.75);
\draw(4.25,0)--(5.5,-.75);
\draw(5.5,.75)node{$\times$};
\draw(5.5,-.75)node{$\times$};
\draw(4.25,-1)node{\scalebox{.8}[.8]{$\lam^1$}};
\draw(6.25,0)node{$+$};
\draw[snake=snake,segment length=4pt](7,0)--(8.25,0);
\draw(8.25,0)--(9.5,-.75);
\draw(9.5,-.75)node{$\times$};
\draw(8.25,0)--(9.15,.54);
\draw(9.25,.6)circle(.1);
\draw(9.35,.66)--(10.25,1.2);
\draw(10.25,1.2)node{$\times$};
\draw(8.5,-1)node{\scalebox{.8}[.8]{$\lam^2$}};
\draw(11.5,0)node{$+\;\;\;\;\cdots$};
\end{tikzpicture}

\noindent and

\begin{tikzpicture}[segment amplitude=2pt]
\draw(0,-.05)node{$y(\eta)\;\;\;=$};
\draw(1.25,0)--(2.25,0);
\draw(2.25,0)node{$\times$};
\draw(1.75,-1.35)node{\scalebox{.8}[.8]{$\lam^0$}};
\draw(3,0)node{$+$};
\draw(3.75,0)--(4.4,0);
\draw(4.5,0)circle(.1);
\draw(4.6,0)--(5.25,0);
\draw(5.25,0)node{$\times$};
\draw(4.5,-1.35)node{\scalebox{.8}[.8]{$\lam^1$}};
\draw(6,0)node{$+$};
\draw(6.75,0)--(7.4,0);
\draw(7.5,0)circle(.1);
\draw(7.6,0)--(8.15,0);
\draw(8.25,0)circle(.1);
\draw(8.35,0)--(9,0);
\draw(9,0)node{$\times$};
\draw(9.75,0)node{$+$};
\draw(10.5,0)--(11.5,0);
\draw(11.5,0)--(12.5,-.6);
\draw(12.5,-.6)node{$\times$};
\draw[snake=snake,segment length=4pt](11.5,0)--(12,.75);
\draw(12,.75)--(13,1.25);
\draw(12,.75)--(13,.25);
\draw(13,1.25)node{$\times$};
\draw(13,.25)node{$\times$};
\draw[snake=brace,mirror snake,segment amplitude=6pt](7,-.9)--(12.75,-.9);
\draw(9.9,-1.35)node{\scalebox{.8}[.8]{$\lam^2$}};
\draw(14.25,0)node{$+\;\;\;\cdots\;$.};
\end{tikzpicture}

\noindent
Here a circle means an insertion of the constant term $x_\cl$, a cross means
attaching to the external noise $\eta$, a wiggly line means the propagator
with the Green's function $G^M$ of $x$, and a solid line means the propagator
with the Green's function $G^m$ of $y$.
To each trivalent vertex with one wiggly line and two solid lines is assigned
the coupling constant $-\lam$ and to each circle on a solid line is assigned
the factor $-\lam x_\cl$.
As usual, each diagramme is divided by the order of its automorphism group.
Note that the quantity represented by each diagramme, though $\tau$-dependent,
is determined by solely the equations in \eqref{L22} and does not rely on the
initial/boundary conditions of a particular solution.

We can express $\bra x(\tau)\ket$ and $\bra y(\tau_1)y(\tau_2)\ket$ by
diagrammes.
Since the same method of iteration is used both here and in Section~\ref{01},
the results to the order $\lam^1$ must agree.
To the order $\lam^2$, we have

\begin{tikzpicture}[segment amplitude=2pt]
\draw(0,0)node{$\bra x(\tau)\ket\;\;\;\;=$};
\draw(1.75,0)circle(.1);
\draw(1.75,.25)node{\scalebox{.8}[.8]{$\tau$}};
\draw(1.75,-1)node{($a$)};
\draw(1.75,-1.5)node{\scalebox{.8}[.8]{$\lam^0$}};
\draw(2.5,0)node{$+$};
\draw[snake=snake,segment length=4pt](3.25,0)--(4.5,0);
\draw(3.25,.25)node{\scalebox{.8}[.8]{$\tau$}};
\draw(4.4,.3)node{\scalebox{.8}[.8]{$\tau'$}};
\draw(5.125,0)circle(.625);
\draw(5.75,0)node{$\times$};
\draw(5.9,.3)node{\scalebox{.8}[.8]{$\tau''$}};
\draw(4.5,-1)node{($b$)};
\draw(4.5,-1.5)node{\scalebox{.8}[.8]{$\lam^1$}};
\draw(6.5,0)node{$+$};
\draw[snake=snake,segment length=4pt](7.25,0)--(8.5,0);
\draw(7.25,.25)node{\scalebox{.8}[.8]{$\tau$}};
\draw(8.4,.3)node{\scalebox{.8}[.8]{$\tau'$}};
\draw(9.35,-.75)arc(-82.5:262.5:.75);
\draw(9.25,.75)node{$\times$};
\draw(9.25,-.75)circle(.1);
\draw(9.45,-.5)node{\scalebox{.8}[.8]{$\tau''$}};
\draw(9.4,1)node{\scalebox{.8}[.8]{$\tau'''$}};
\draw(8.5,-1)node{($c$)};
\draw(8.5,-1.5)node{\scalebox{.8}[.8]{$\lam^2$}};
\draw(11.25,0)node{$+\;\;\;\;\cdots$};
\end{tikzpicture}

\noindent and

\begin{tikzpicture}[segment amplitude=2pt]
\draw(0,.5)node{$\bra y(\tau_1)y(\tau_2)\ket\;\;\;=$};
\draw(2,.5)--(4,.5);
\draw(3,.5)node{$\times$};
\draw(2,.72)node{\scalebox{.8}[.8]{$\tau_2$}};
\draw(4,.72)node{\scalebox{.8}[.8]{$\tau_1$}};
\draw(3,.8)node{\scalebox{.8}[.8]{$\tau'$}};
\draw(3,0)node{($d$)};
\draw(3,-.6)node{\scalebox{.8}[.8]{$\lam^0$}};
\draw(4.75,.5)node{$+$};
\draw(5.5,.5)--(7.4,.5);
\draw(6.5,.5)node{$\times$};
\draw(7.5,.5)circle(.1);
\draw(7.6,.5)--(8.5,.5);
\draw(5.5,.72)node{\scalebox{.8}[.8]{$\tau_2$}};
\draw(8.5,.72)node{\scalebox{.8}[.8]{$\tau_1$}};
\draw(6.5,.8)node{\scalebox{.8}[.8]{$\tau'_2$}};
\draw(7.5,.8)node{\scalebox{.8}[.8]{$\tau'$}};
\draw(7,0)node{($e$)};
\draw(9.25,.5)node{$+$};
\draw(10,.5)--(10.9,.5);
\draw(11,.5)circle(.1);
\draw(11.1,.5)--(13,.5);
\draw(12,.5)node{$\times$};
\draw(10,.72)node{\scalebox{.8}[.8]{$\tau_2$}};
\draw(13,.72)node{\scalebox{.8}[.8]{$\tau_1$}};
\draw(11,.8)node{\scalebox{.8}[.8]{$\tau'$}};
\draw(12,.8)node{\scalebox{.8}[.8]{$\tau'_1$}};
\draw(11.5,0)node{($f$)};
\draw[snake=brace,mirror snake,segment amplitude=6pt](6,-.25)--(12.5,-.25);
\draw(9.25,-.6)node{\scalebox{.8}[.8]{$\lam^1$}};
\draw(13.75,.5)node{$+$};
\draw(-1,-3)node{$+$};
\draw(-.5,-2)--(.15,-2);
\draw(.25,-2)circle(.1);
\draw(.35,-2)--(1.65,-2);
\draw(1,-2)node{$\times$};
\draw(1.75,-2)circle(.1);
\draw(1.85,-2)--(2.5,-2);
\draw(-.5,-1.78)node{\scalebox{.8}[.8]{$\tau_2$}};
\draw(2.5,-1.78)node{\scalebox{.8}[.8]{$\tau_1$}};
\draw(.25,-1.7)node{\scalebox{.8}[.8]{$\tau'_2$}};
\draw(1.75,-1.7)node{\scalebox{.8}[.8]{$\tau'_1$}};
\draw(1,-1.7)node{\scalebox{.8}[.8]{$\tau'$}};
\draw(3,-2)node{($g$)};
\draw(-.5,-3)--(.15,-3);
\draw(.25,-3)circle(.1);
\draw(.35,-3)--(.9,-3);
\draw(1,-3)circle(.1);
\draw(1.1,-3)--(2.5,-3);
\draw(1.75,-3)node{$\times$};
\draw(-.5,-2.78)node{\scalebox{.8}[.8]{$\tau_2$}};
\draw(2.5,-2.78)node{\scalebox{.8}[.8]{$\tau_1$}};
\draw(.25,-2.7)node{\scalebox{.8}[.8]{$\tau'_2$}};
\draw(1,-2.7)node{\scalebox{.8}[.8]{$\tau''_2$}};
\draw(1.75,-2.7)node{\scalebox{.8}[.8]{$\tau'$}};
\draw(3,-3)node{($h$)};
\draw(-.5,-4)--(.9,-4);
\draw(.25,-4)node{$\times$};
\draw(1,-4)circle(.1);
\draw(1.1,-4)--(1.65,-4);
\draw(1.75,-4)circle(.1);
\draw(1.85,-4)--(2.5,-4);
\draw(-.5,-3.78)node{\scalebox{.8}[.8]{$\tau_2$}};
\draw(2.5,-3.78)node{\scalebox{.8}[.8]{$\tau_1$}};
\draw(.25,-3.7)node{\scalebox{.8}[.8]{$\tau'$}};
\draw(1,-3.7)node{\scalebox{.8}[.8]{$\tau''_1$}};
\draw(1.75,-3.7)node{\scalebox{.8}[.8]{$\tau'_1$}};
\draw(3,-4)node{($i$)};
\draw(3.75,-3)node{$+$};
\draw(4.25,-2)--(5,-2);
\draw[snake=snake,segment length=4pt](5,-2)--(6.5,-2);
\draw(5,-2)arc(-180:0:.75);
\draw(5.75,-2.75)node{$\times$};
\draw(6.5,-2)--(8,-2);
\draw(7.25,-2)node{$\times$};
\draw(4.25,-1.78)node{\scalebox{.8}[.8]{$\tau_2$}};
\draw(8,-1.78)node{\scalebox{.8}[.8]{$\tau_1$}};
\draw(5,-1.7)node{\scalebox{.8}[.8]{$\tau'_2$}};
\draw(6.5,-1.7)node{\scalebox{.8}[.8]{$\tau'_1$}};
\draw(7.25,-1.7)node{\scalebox{.8}[.8]{$\tau'$}};
\draw(5.75,-2.45)node{\scalebox{.8}[.8]{$\tau''$}};
\draw(8.5,-2)node{($j$)};
\draw(4.25,-3.5)--(5.75,-3.5);
\draw(5,-3.5)node{$\times$};
\draw[snake=snake,segment length=4pt](5.75,-3.5)--(7.25,-3.5);
\draw(7.25,-3.5)--(8,-3.5);
\draw(5.75,-3.5)arc(-180:0:.75);
\draw(6.5,-4.25)node{$\times$};
\draw(4.25,-3.28)node{\scalebox{.8}[.8]{$\tau_2$}};
\draw(8,-3.28)node{\scalebox{.8}[.8]{$\tau_1$}};
\draw(5.75,-3.2)node{\scalebox{.8}[.8]{$\tau'_2$}};
\draw(7.25,-3.2)node{\scalebox{.8}[.8]{$\tau'_1$}};
\draw(5,-3.2)node{\scalebox{.8}[.8]{$\tau'$}};
\draw(6.5,-3.95)node{\scalebox{.8}[.8]{$\tau''$}};
\draw(8.5,-3.5)node{($k$)};
\draw(9.25,-3)node{$+$};
\draw(9.75,-2.5)--(12,-2.5);
\draw[snake=snake,segment length=4pt](10.5,-2.5)--(10.5,-2);
\draw(10.5,-1)node{$\times$};
\draw(10.5,-1.5)circle(.5);
\draw(11.25,-2.5)node{$\times$};
\draw(9.75,-2.3)node{\scalebox{.8}[.8]{$\tau_2$}};
\draw(12,-2.3)node{\scalebox{.8}[.8]{$\tau_1$}};
\draw(11.25,-2.2)node{\scalebox{.8}[.8]{$\tau'$}};
\draw(10.5,-2.7)node{\scalebox{.8}[.8]{$\tau'_2$}};
\draw(10.5,-1.75)node{\scalebox{.8}[.8]{$\tau''_2$}};
\draw(10.5,-.75)node{\scalebox{.8}[.8]{$\tau''$}};
\draw(12.5,-2.5)node{($l$)};
\draw(9.75,-4.75)--(12,-4.75);
\draw(10.5,-4.75)node{$\times$};
\draw[snake=snake,segment length=4pt](11.25,-4.75)--(11.25,-4.25);
\draw(11.25,-3.75)circle(.5);
\draw(11.25,-3.25)node{$\times$};
\draw(9.75,-4.55)node{\scalebox{.8}[.8]{$\tau_2$}};
\draw(12,-4.55)node{\scalebox{.8}[.8]{$\tau_1$}};
\draw(10.5,-4.45)node{\scalebox{.8}[.8]{$\tau'$}};
\draw(11.25,-4.95)node{\scalebox{.8}[.8]{$\tau'_1$}};
\draw(11.25,-4)node{\scalebox{.8}[.8]{$\tau''_1$}};
\draw(11.25,-3)node{\scalebox{.8}[.8]{$\tau''$}};
\draw(12.5,-4.75)node{($m$)};
\draw[snake=brace,mirror snake,segment amplitude=6pt](-.5,-5.25)--(12.5,-5.25);
\draw(6,-5.65)node{\scalebox{.8}[.8]{$\lam^2$}};
\draw(13.75,-3)node{$+\;\;\;\cdots\;$,};
\end{tikzpicture}

\noindent
where a cross on a solid line means now contraction of two $\eta$'s in the
diagrammatic expansions of $x(\eta)$ and $y(\eta)$ above using \eqref{eta^2},
giving rise to a factor of $2b$.
Quantitatively, these diagrammes are
\begin{align*}
(a)&=x_\cl,    \\
(b)&=-b\lam\mint0\tau\d\tau'\,G^M(\tau-\tau')
\mint0{\tau'}\d\tau''\,G^m(\tau'-\tau'')^2, \\
(c)&=2b\lam^2 x_\cl\,\mint0\tau\d\tau'\,G^M(\tau-\tau')
\mint0{\tau'}\d\tau''\,G^m(\tau'-\tau'')
\mint{\tau''}{\tau'}\d\tau'''\,G^m(\tau'-\tau''')G^m(\tau'''-\tau'')
\end{align*}
for $\bra x(\tau)\ket$.
The diagrammes ($a$) and ($b$) agree with \eqref{<x>} when $\tau\to+\infty$.
For $\bra y(\tau_1)y(\tau_2)\ket$, we have
\begin{align*}
(d)&=2b\mint0{\min(\tau_1,\tau_2)}\d\tau'\,G^m(\tau_1-\tau')G^m(\tau_2-\tau'),
\\
(e)&=-2b\lam x_\cl\,\mint0{\min(\tau_1,\tau_2)}\d\tau'\,G^m(\tau_1-\tau')
\mint{\tau'}{\tau_2}\d\tau_2'\,G^m(\tau_2-\tau_2')G^m(\tau_2'-\tau'), \\
(f)&=-2b\lam x_\cl\,\mint0{\min(\tau_1,\tau_2)}\d\tau'\,G^m(\tau_2-\tau')
\mint{\tau'}{\tau_1}\d\tau'_2\,G^m(\tau_1-\tau_1')G^m(\tau_1'-\tau'), \\
(g)&=2b\lam^2x_\cl^2\,\mint0{\tau_1}\d\tau'_1\,G^m(\tau_1-\tau_1')
\mint0{\tau_2}\d\tau'_2\,G^m(\tau_2-\tau_2')
\mint0{\min(\tau'_1,\tau'_2)}\d\tau'\,G^m(\tau'_1-\tau')G^m(\tau'_2-\tau'),
\\
(h)&=2b\lam^2x_\cl^2\,\mint0{\min(\tau_1,\tau_2)}\d\tau'\,G^m(\tau_1-\tau')
\mint{}{}\!\!\mint{\tau'\le\tau''_2\le\tau'_2\le\tau_2}{}\d\tau'_2\d\tau''_2\,
G^m(\tau_2-\tau'_2)G^m(\tau'_2-\tau''_2)G^m(\tau''_2-\tau'), \\
(i)&=2b\lam^2x_\cl^2\,\mint0{\min(\tau_1,\tau_2)}\d\tau'\,G^m(\tau_2-\tau')
\mint{}{}\!\!\mint{\tau'\le\tau''_1\le\tau'_1\le\tau_1}{}\d\tau'_1\d\tau''_1\,
G^m(\tau_1-\tau'_1)G^m(\tau'_1-\tau''_1)G^m(\tau''_1-\tau'), \\
(j)&=4b^2\lam^2\mint0{\min(\tau_1,\tau_2)}\d\tau'\,G^m(\tau_1-\tau')
\mint{}{}\!\!\mint{\tau'\le\tau'_1\le\tau'_2\le\tau_2}{}\d\tau'_1\d\tau'_2
\,G^m(\tau'_1-\tau')G^m(\tau_2-\tau'_2)G^M(\tau'_1-\tau'_2) \\
&\qquad\mint{\tau'_1}{\tau'_2}\d\tau''\,G^m(\tau'_2-\tau'')G^m(\tau''-\tau'_1)
     \\
(k)&=4b^2\lam^2\mint0{\min(\tau_1,\tau_2)}\d\tau'\,G^m(\tau_2-\tau')
\mint{}{}\!\!\mint{\tau'\le\tau'_2\le\tau'_1\le\tau_1}{}\d\tau'_1\d\tau'_2
\,G^m(\tau'_2-\tau')G^m(\tau_1-\tau'_1)G^M(\tau'_2-\tau'_1) \\
&\qquad\mint{\tau'_2}{\tau'_1}\d\tau''\,G^m(\tau'_1-\tau'')G^m(\tau''-\tau'_2)
     \\
(l)&=4b^2\lam^2\mint0{\min(\tau_1,\tau_2)}\d\tau'\,G^m(\tau_1-\tau')
\mint{\tau'}{\tau_2}\d\tau'_2\,G^m(\tau_2-\tau'_2)G^m(\tau'_2-\tau') \\
&\qquad\mint0{\tau'_2}\d\tau''_2\,G^M(\tau'_2-\tau''_2)
\mint0{\tau''_2}\d\tau''\,G^m(\tau''_2-\tau'')^2  \\
(m)&=4b^2\lam^2\mint0{\min(\tau_1,\tau_2)}\d\tau'\,G^m(\tau_2-\tau')
\mint{\tau'}{\tau_1}\d\tau'_1\,G^m(\tau_1-\tau'_1)G^m(\tau'_1-\tau') \\
&\qquad\mint0{\tau'_1}\d\tau''_1\,G^M(\tau'_1-\tau''_1)
\mint0{\tau''_1}\d\tau''\,G^m(\tau''_1-\tau'')^2,
\end{align*}
and diagrammes ($d$), ($e$), ($f$) agree with \eqref{<y^2>} when
$\tau_1=\tau_2=\tau\to+\infty$.

\section{Supersymmetric representation and a possible Fokker-Planck
Hamiltonian}
Independent of the goals of obtaining a possible definition of quantum gravity
and (ambitious) physical predictions as outlined just above, it was suggested
in \cite{cosmob} that, by using the idea of stochastic time with a second order
time evolution, a $d$-dimensional Euclidean QFT can be defined as the limit
of a $(d+1)$-dimensional QFT with a double Fock space structure.
The latter contains the usual Fock space for the quantum particle degrees of
freedom and another Fock space for the accessible levels of what we find
suggestive to name the ``stochastic time dependent oscillating vacuum (STDOV).
This builds a good theoretical framework for the quantum gravity scenario
in~\cite{cosmob}.
A higher order Langevin equation could also reveal some properties of simpler
quantum mechanical models, such as the one with a conformal potential
$\sim1/x^2$, a system which never reaches an equilibrium but has interesting
properties at a finite stochastic time.

This $(d+1)$-dimensional QFT is Euclidean in all coordinates but the
stochastic time, and it involves effectively a physical UV cutoff
$1/\Del T$ as a parameter that occurs naturally from dimensionality
considerations in a second order Langevin equation.

The correlation functions stemming from a Langevin equation can be represented
by a supersymmetric path integral, with the generating functional
\begin{align}
Z[J_x,J_y]=&\mint{}{}[\d x]_\tau[\d y]_\tau[\d\psi_x]_\tau[\d\psi_y]_\tau
[\d\bar\psi_x]_\tau[\d\bar\psi_y]_\tau[\d\eta_x]_\tau[\d\eta_y]_\tau \nn\\
&\exp\Big[\!-\!\sfrac{1}{\hb}\mint0T\d\tau\,\big({\cal L}_\susy(x,y,\psi_x,
\psi_y,\bar\psi_x,\bar\psi_y,\eta_x,\eta_y)+xJ_x+yJ_y\big)\Big]
\end{align}
using the standard determinant manipulation \cite{gozzi}.
For simplicity, we will write $q$ for $(x,y)$, $\eta$ for $(\eta_x,\eta_y)$,
$\psi$ for $(\psi_x,\psi_y)$, and $\bar\psi$ for $(\bar\psi_x,\bar\psi_y)$.
The action is topological in the sense that ${\cal L}_\susy =s_\topo(\cdots)$,
where the supersymmetry transformations
\bq\label{Qsusy}
s_\topo q=\psi,\quad s_\topo\psi=0,\quad s_\topo\bar\psi=\eta
\eq
satisfy $s_\topo^2=0$.
In our case with a second order time evolution, one has
\begin{align}\label{Lsusy}
{\cal L}_\susy&=s_\topo\big(\bar\psi\big(a^2\ddot q+2b\,\dot q
+\sfrac{\pa S}{\pa q}-\sfrac12\eta\big)\big)  \nn\\
&=-\sfrac12\eta^2+\eta\big(a^2\ddot q+2b\,\dot q
+\sfrac{\pa S}{\pa q}\big)-\bar\psi\big(a^2\ddot\psi+2b\,\dot\psi
+\sfrac{\pa^2 S}{\pa q^2}\psi\big)  \nn\\
&\sim\sfrac12\big(a^2\ddot q+2b\,\dot q+\sfrac{\pa S}{\pa q}\big)^2
-\bar\psi\big(a^2\ddot\psi+2b\,\dot\psi+\sfrac{\pa^2 S}{\pa q^2}\psi\big),
\end{align}
This Lagrangian ${\cal L}_\susy$ was mentioned in Eq.~\eqref{mention} and
can be called the Fokker-Planck supersymmetric Lagrangian.

The existence of a supersymmetric action that one can associate to an
evolution equation involving a Gaussian noise, not necessarily a white one,
adding to the drift force, is a general property \cite{parisi}.
This supersymmetry has been investigated with care in \cite{gozzi} in the
context of stochastic quantisation.
The relation between this fact and topological quantum field theory was noted
in \cite{stoctqft}.

When one eliminates the fermions in the path integral by Berezin integration
and computes correlation functions, one recovers the same results as from
solving the Langevin equations and computing mean values of functions of $q$
by elimination of the noise $\eta$.

It is actually useful to express by Legendre transform the Hamiltonian
associated to the supersymmetric Lagrangian~\eqref{Lsusy}.
Let us recall that, when $a=0$, $b=\frac12$, $\Ps$ and $\bar\Ps$ are
self-conjugate.
In this case, the supersymmetric Hamiltonian can be represented as a
$2\times2$ matrix whose diagonal terms are both advanced and retarded
Fokker-Planck Hamiltonians for the first order stochastic time evolution,
as they can be computed in statistical mechanics.
The property of its spectrum determines the convergence of the stochastic
process~\cite{gozzi}.
In this sense, the supersymmetry is truly a topological symmetry that
determines the Fokker-Planck process from first principles.
The word ``topological'' is justified by the fact that many interesting
theories express in the simplest way their physics only in the slice
$\tau=\infty$ \cite{stoctqft}.\footnote{In the general case of an action in
$d$ dimensions $S_d$, the $d+1$ dimensional supersymmetric stochastic
Lagrangian that computes the correlators in the bulk is denoted by
${\cal L}_\susy^{d+1}$.
Its action is basically $s_\topo$-exact under a nilpotent operator, as in
eq.~\eqref{Qsusy}.
One can compute its supersymmetric Hamiltonian for the stochastic time
evolution that we denote as ${\cal H}_d$ for future use in the text.
We express, without proof, the fact that the spectrum of ${\cal H}_d$ is made
of (bound) states evolving in $d+1$ dimensions, with $d$-dimensional wave
functions related to the instantons of $S_d$.}

For $a\neq0$, we have a more general supersymmetric Lagrangian that has
higher order derivative terms, symbolically $L=L(q,\dot q,\ddot q)$.
This doesn't change our point of view that its Hamiltonian, once suitably
defined, will generate the stochastic time evolution, and define the
required generalisation of the Fokker-Plank evolution kernel.

To see how it works in an elementary way, we can use the simple
$\eta$-dependence of the Lagrangian.
We can write the bosonic part of the Lagrangian ${\cal L}_\susy$ as
\bq
{\cal L}^B_\susy=-a^2\dot\eta\dot q+b(\eta\dot q-q\dot\eta)-\sfrac12\eta^2
+\eta\,\sfrac{\pa S}{\pa q}
\eq
up to a total derivative.
Then the canonical momenta of $q$ and $\eta$ are
\bq
p_q\equiv\sfrac{\pa{\cal L}^B_\susy}{\pa\dot q}=-a^2\dot\eta+b\,\eta,\qquad
p_\eta\equiv\sfrac{\pa{\cal L}^B_\susy}{\pa\dot\eta}=-a^2\dot q-b\,q,
\eq
respectively.
Taking a Legendre transform, the bosonic part of the Hamiltonian is
\begin{align}\label{Hsusy}
{\cal H}^B_\susy&=p_q\,\dot q+p_\eta\,\dot\eta-{\cal L}^B_\susy \nn\\
&=-\sfrac{1}{a^2}(p_q-b\,\eta)(p_\eta+b\,q)
+\sfrac12\,\eta^2-\eta\,\sfrac{\pa S}{\pa q}.
\end{align}
It shows that we must consider a doubling of the phase space,
$(q,p_q)\to(q,p_q,\eta,p_\eta)$, and that at finite values of $\tau$, the
limit $a\to0$ is not smooth.
The fermionic part contains the supersymmetric partners and has the same
structure.
This extension of the phase space is because the Langevin equation is of a
higher order.\footnote{An analogous phenomenon arises when one performs the
BRST quantisation of supergravity, where the fermionic Lagrange multiplier
for the gauge-fixing of the Rarita-Schwinger field becomes a propagating
fermionic field.}

If we examine the Hamiltonian \eqref{Hsusy} from a perturbative point of view,
its zeroth order is given by a quadratic approximation of $S$ as a function
of $x$.
If we restrict to this approximation, one can diagonalise ${\cal H}^B_\susy$
in the space $(x,p_x)\to (x,p_x,\eta,p_\eta)$ and get a double harmonic
oscillator with a double harmonic spectrum, one for the particles and one
for the noise through the quantisation of $\eta$.
In our approximation were we neglect $\eta_x$, this extra quantisation is
reported on the oscillations at finite time of $x$ around $x_\cl$, till it
becomes stationary.
These features survive perturbatively for a large class of potentials,
and one may hope that they can be true non-perturbatively.

In a different approach, and basically with the same conclusion, we can
analyse the Lagrangian ${\cal L}_\susy$ directly after the elimination of
$\eta$, which gives the last line in~\eqref{Lsusy}.
The Lagrangian ${\cal L}_\susy$ has higher order derivatives and can be
studied by applying the standard Lagrangian/Hamiltonian formalism of
Ostrogradsky (see for example \cite{highercan} and references
therein).\footnote{The same remark applies to the models presented in
\cite{stoctqft}.}
Given a Lagrangian depending on $q,\dot q,\ddot q$, where the dot means
$\dot\,\equiv\frac{\d}{\d\tau}$, and an action
$S=\int\d\tau\,L(q,\dot q,\ddot q)$, one finds that the extrema of $S$ occur
when the Euler-Lagrange equation of motion
\bq
\mfrac{\pa L}{\pa q}=\mfrac{\d}{\d\tau}\mfrac{\pa L}{\pa\dot q}
-\mfrac{\d^2}{\d\tau^2}\mfrac{\pa L}{\pa\ddot q}
\eq
is satisfied.
In fact, a general variation of $L$ for arbitrary variations $\del q$ and
$\del\dot q$ is
\bq\label{var}
\del L=\Big(\mfrac{\pa L}{\pa q}-\mfrac{\d}{\d\tau}\mfrac{\pa L}{\pa\dot q}
+\mfrac{\d^2}{\d\tau^2}\mfrac{\pa L}{\pa\dot q}\Big)\,\del q
+\mfrac{\d}{\d\tau}\,(p_0\del q_0 +p_1\del\dot q_1),
\eq
where
\bq\label{phasespace}
q_0\equiv q,\qquad q_1\equiv\dot q,\qquad
p_0\equiv\mfrac{\pa L}{\pa\dot q}-\mfrac{\d}{\d\tau}\mfrac{\pa L}{\pa\ddot q},
\qquad p_1\equiv\mfrac{\pa L}{\pa\ddot q}.
\eq
This shows that the phase space is parametrised by the conjugate coordinates
$(q_0,q_1,p_0,p_1)$.
The canonical phase space is doubled, as can be simply understood because
one needs twice as many initial conditions. {The Ostrogradsky
formalism is valid for higher order Lagrangians
$L(q,\dot q,\ddot q,\dddot q,\cdots,q^{(r)}))$, giving an enlarged phase
space of dimension $2r$.}

This doubling of the phase space has non-trivial consequences.
In particular, when $a\neq 0$, $\dot q$ is not identified as the momentum of
$q$ in the Lagrangian \eqref{Lsusy}, so $q$ and $\dot q$ can be measured
simultaneously in the quantum mechanics defined by replacing the
(anti)coordinates by operators and the Poisson brackets by (anti)commutators.

In fact, the uncertainty relation holds only between $q_0$ and $p_0$
(for $a\neq 0$), and between $q_1$ and $p_1$ (see the canonical commutation
relations below).

Eq.~\eqref{var} shows that the conserved Hamiltonian associated to
$L(q,\dot q,\ddot q)$, which expresses the $\tau$-evolution, is still given
by a Legendre transformation
\bq
H\equiv p_0\dot q_0+p_1\dot q_1-L(q,\dot q,\ddot q),
\eq
where $\dot q, \ddot q$ must be expressed as functions of $q_0,p_0,q_1,p_1$
using \eqref{phasespace}.
One finds that the phase space equations of motion are
\[ \dot p_0=-\mfrac{\pa H}{\pa q_0},\quad\dot p_1=-\mfrac{\pa H}{\pa q_1},
\quad\dot q_0=\mfrac{\pa H}{\pa p_1},\quad\dot q_1=\mfrac{\pa H }{\pa p_0} \]
They are first order equations, just as in the standard case, but a doubling.
Quantisation in the operator formalism is then defined by regarding all
coordinates and momenta as operators subject to the canonical relations
\bq
[\hat q_i,\hat p_j]=i\hbar\,\del_{ij},\quad
[\hat q_i,\hat q_j]=[\hat p_i,\hat p_j]=0.
\eq
and for any operator $\hat A$, one has $\dot{\hat A}=[\hat H,\hat A]$.
This more systematic construction explains in a different way the doubling of
the Fock space that we directly derived, before the elimination of the noises.
The positivity of the spectrum is not ensured, as it was obvious seen in
\eqref{Hsusy}, and its physical interpretation is different than of standard
quantum mechanics in its Euclidean form.
As explained in this paper, the later can be only obtained in limit
$\tau\to\infty$, if the limit exists.

The construction of the phase space can be easily repeated for the entire
supersymmetric ${\cal L}_\susy$ in Eq.~\eqref{Lsusy}, giving a graded
symplectic structure and a supersymmetric Hamiltonian, where both the $\Ps$
and $\bar\Ps$ have their own independent momenta for $a\neq 0$ and there is
also a doubling for the fermionic part of the phase space as compared to the
case $a=0$.

All this suggests that a well-defined Fokker-Planck Hamiltonian, which defines
the stochastic time evolution by first order Hamiltonian equations, can be
associated to the second order Langevin equation.
This explains in a better way the expression \eqref{Hsusy}, where the noise
was kept in a rather artificial way.
The supersymmetric representation \eqref{Lsusy} of the second order Langevin
equation is expected to play a key role in the understanding of the
$\tau$-evolution as it did in the case of a first order Langevin equation.
It actually used in a separate reference where we analyse the covariance
questions of the gravity stochastic equation.
Appendix~B describes this for the example given in Appendix~A.

\end{document}